\documentclass[journal,a4paper,12pt,onecolumn,draftclsnofoot]{IEEEtran}
\usepackage[pdftex]{graphicx}

\usepackage{nidanfloat}

\usepackage{flushend}

\usepackage{multirow}
\usepackage{amssymb, bm, setspace, comment}

\usepackage{cite}

\ifCLASSINFOpdf
\else
\fi
\usepackage{epsfig}

\usepackage[cmex10]{amsmath}
\makeatletter
\let\MYcaption\@makecaption
\makeatother

\usepackage[skip=1pt]{subcaption}

\makeatletter
\let\@makecaption\MYcaption
\makeatother

\usepackage{cite}
\usepackage{amssymb}
\usepackage{cite}
\graphicspath{{./images/}}
\usepackage{bm}
\usepackage{multirow}

\usepackage{here}

\begin{document}
%

\title{Higher-order tensor independent component analysis to realize MIMO remote sensing of respiration and heartbeat signals}

%
%

\author{Seishiro~Goto,~Ryo~Natsuaki and~Akira~Hirose
\thanks{A part of this work was supported by JSPS KAKENHI Grant No.18H04105 and also by the Cooperative Research Project Program of the Research Institute of Electrical Communication (RIEC), Tohoku University.}%
\thanks{S.~Goto is with the Department of Bioengineering, The University of Tokyo, Tokyo 113-8656, Japan.
        \tt\small{ 
        seishiro\_goto@eis.t.u-tokyo.ac.jp}}%
\thanks{R.~Natsuaki is with the Department of Electrical Engineering and Information Systems, The University of Tokyo, Tokyo 113-8656, Japan.
        \tt\small{
        natsuaki@ee.t.utokyo.ac.jp}}%
\thanks{A.~Hirose is with the Department of Bioengineering and also with the Department of Electrical Engineering and Information Systems, The University of Tokyo, Tokyo 113-8656, Japan.
        \tt\small{ 
        ahirose@ee.t.u-tokyo.ac.jp}}%
}
\maketitle


%
\begin{abstract}
This paper proposes a novel method of independent component analysis (ICA), which we name higher-order tensor ICA (HOT-ICA). HOT-ICA is a tensor ICA that makes effective use of the signal categories represented by the axes of a separating tensor. Conventional tensor ICAs, such as multilinear ICA (MICA) based on Tucker decomposition, do not fully utilize the high dimensionality of tensors because the matricization in MICA nullifies the tensor axial categorization. In this paper, we deal with multiple-target signal separation in a multiple-input multiple-output (MIMO) radar system to detect respiration and heartbeat. HOT-ICA realizes high robustness in learning by incorporating path information, i.e., the physical-measurement categories on which transmitting/receiving antennas were used. In numerical-physical experiments, our HOT-ICA system effectively separate the bio-signals successfully even in an obstacle-affecting environment, which is usually a difficult task. The results demonstrate the significance of the HOT-ICA, which keeps the tensor categorization unchanged for full utilization of the high-dimensionality of the separation tensor.
\end{abstract}


%
\IEEEpeerreviewmaketitle

%
%
%
%



\section{Introduction}
\label{s:introduction}

Conventional heartbeat sensing systems use contact-type electrodes put on a human body. However, recent vital sign detectors sometimes employ noncontact methods. After the first report of respiration detection using microwave \cite{lin1975noninvasive}, there have been a lot of research on respiration and heartbeat measurement based on Doppler radar. Some of them assumed a line-of-sight situation \cite{li2006experiment,li2008random,gu2010instrument,huang2016self}, while others included obstacles such as rubble for measurement in disasters \cite{chen2000microwave,arai2001survivor,wang2018through,donelli2011rescue,bezer2016proposal,igarss}.

Multiple-input multiple-output (MIMO) configuration using multiple transmitting and receiving antennas holds the ability of target and/or path identification. For example, a 24~GHz frequency-modulation continuous-wave (FMCW) MIMO radar detects respiration and heartbeat information for respective targets by focusing on the target distances to separate the individuals \cite{sakuratech}. However, the use of such a high frequency limits its practical applications only within short-range line-of-sight situations. A lower-frequency continuous-wave (CW) radar system has a potential to realize target detection with a wide sensitive area even including obstacles.

Environments including obstacles and multiple targets often require separation of a target signal from others and noise.
A signal-source separation experiment was reported \cite{Sakamoto2018} in which the heartbeat signal of a target was separated from that of another one by using beamforming in an X-band array radar.
However, it is desired to use lower-frequency microwaves in an environment where obstacles exist. Microwave is capable of propagating among obstructions though its range resolution is usually low.
In such a case, a blind source separation (BSS) is promising.
BSS is a framework to estimate original signals from multiple mixed signals based on information itself.
Independent component analysis (ICA) is a typical method in BSS.
ICA removes noise and/or separate targets by finding a separation matrix to linearly transform mixed signals to unmixed ones based on the signals' statistical properties. 
ICA was proposed in 1984 \cite{herault1984circuits}.
Its history is available in Ref.~\cite{jutten2000source}, and the mathematical analyses and many algorithms are described in, e.g., Ref.~\cite{oja2001ica}.
ICA has been frequently dealt with in the field of audio signal processing in the frequency domain \cite{ikeda1999method, sawada2003polar, permutation2004}.

In the radar sensing and imaging field, a system \cite{donelli2011rescue} treated in-phase and the quadrature components obtained by orthogonal detection as two independent real number signals to apply ICA.
However, a pair of in-phase and orthogonal components are essentially a single complex signal to be processed in the framework of complex-valued neural networks \cite{cvnn2012_second,ahirose2012:_gener_chara_of_compl_value_feedf_neura_netwo_in_terms_of_signa_coher,HiroseEckmiller1996IEEENN}.
This paper also deals with complex signals as an entity.
The measurement environment, in addition, changes depending on target movement and obstacle occurrence.
We thus aim to process the properties of complex signals adaptively with time-sequential observation just like the systems in Refs.~\cite{cichocki1994robust} and \cite{cardoso1996equivariant} managing new data every time a data is fed.
The scheme is called online ICA.

Signal processing in measurement using MIMO configuration leads to a construction of data tensor having multiple axes involving path/category information, rather than a data vector representing received signals evenly. A tensor data requires a higher-order tensor for signal separation.
Multilinear ICA (MICA) was proposed for incorporating third-order tensors into the ICA processing.
MICA uses higher-order singular value decomposition (HOSVD) or higher-order orthogonal iteration (HOOI). Their calculation is based on the tensor decomposition proposed by Tucker \cite{tucker1966}. MICA has been positively evaluated for its separation effectiveness\cite{Anh2010,Y_Li2013,Tamara2009,Zhou_Cichocki_2012,SheehanS07hooi}.
First, a core tensor is created by performing singular value decomposition based on matricization of respective modes of the third-order tensor.
A core tensor is equivalent to a diagonal singular value matrix in the conventional singular value decomposition.
Then, an approximate separate tensor is constructed by using the core tensor and a factor matrix generated by singular value decomposition in respective modes of matricization.
As a result, an error tensor is calculated by using the approximation tensor and the original data tensor.
The approximate separation tensor is updated iteratively until the error tensor becomes sufficiently small.

Some MICA methods \cite{Furuhashi2015,cichocki2013tensor,ZhouCichocki2010} treat brain waves as waveform data.
For example, Ref.~\cite{cichocki2013tensor} employed three categories, namely, its frequency bin, time frame, and symptoms, as the axes of the data tensor. It proposed a method to identify illness from patient data by MICA for clustering. MICA is also used frequently in the field of image data analysis \cite{MAOVasi2005,MAOVasi2007}.
Besides, Ref.~\cite{AI2013186} proposed a method that extends MICA to higher dimension.

It is true that the methods such as HOSVD and HOOI based on Tucker decomposition can process data tensors in the framework of MICA as described above.
However, they do not utilize the nature of the higher order tensors effectively. That is, the categories in the data represented by the tensor axes is nullified by the matricization.
However, it should be possible to realize tensor ICA processing more meaningfully in such a manner that the data categories represented by the axes remain undestructed for enhanced functionality.

This paper proposes such a method, namely, higher-order tensor independent component analysis (HOT-ICA), that realizes an effective use of the tensor structure representing data categories such as respective origins of individual data.
We demonstrate its effectiveness in numerical experiments of non-contact respiration and heartbeat detection for multiple people in an environment with obstacles.

We deal with a CW MIMO Doppler radar to show the strength of the HOT-ICA processing.
HOT-ICA is capable of manipulating the contributions of categorical components in the separation tensor in its independence evaluation. For instance, HOT-ICA with sensitivity control in separation-tensor updates will make MIMO remote sensing more robust to obstacles in electromagnetic-wave propagation.
In this paper, we perform a numerical experiment which reflects the instability of received signals due to the environmental changes caused by obstacle insertion.
The numerical experiment demonstrates the effectiveness of HOT-ICA with sensitivity control in the separation tensor updates. We compare the result with a conventional method, namely, complex-valued frequency-domain ICA (CF-ICA) \cite{igarss}.
It is found that the HOT-ICA realizes the essential use of tensors in the ICA.

This paper is organized as follows.
Section \ref{s:proposal} explains the Doppler radar and ICA conventional theory.
Section \ref{sec:onlinehotica} introduces the theory of HOT-ICA proposed in this paper.
Section \ref{s:numerical} describes the system setup and data settings used in the experiment.
Section \ref{s:numericalresult} presents experimental results and discussion.
Section \ref{s:conclusion} is the conclusion of this paper.

\section{Physical and mathematical background}
\label{s:proposal}

\subsection{Doppler radar}
\label{subsec:dopplarradar}

\begin{figure}[H]
  \centering
  \includegraphics[width=0.7\linewidth]{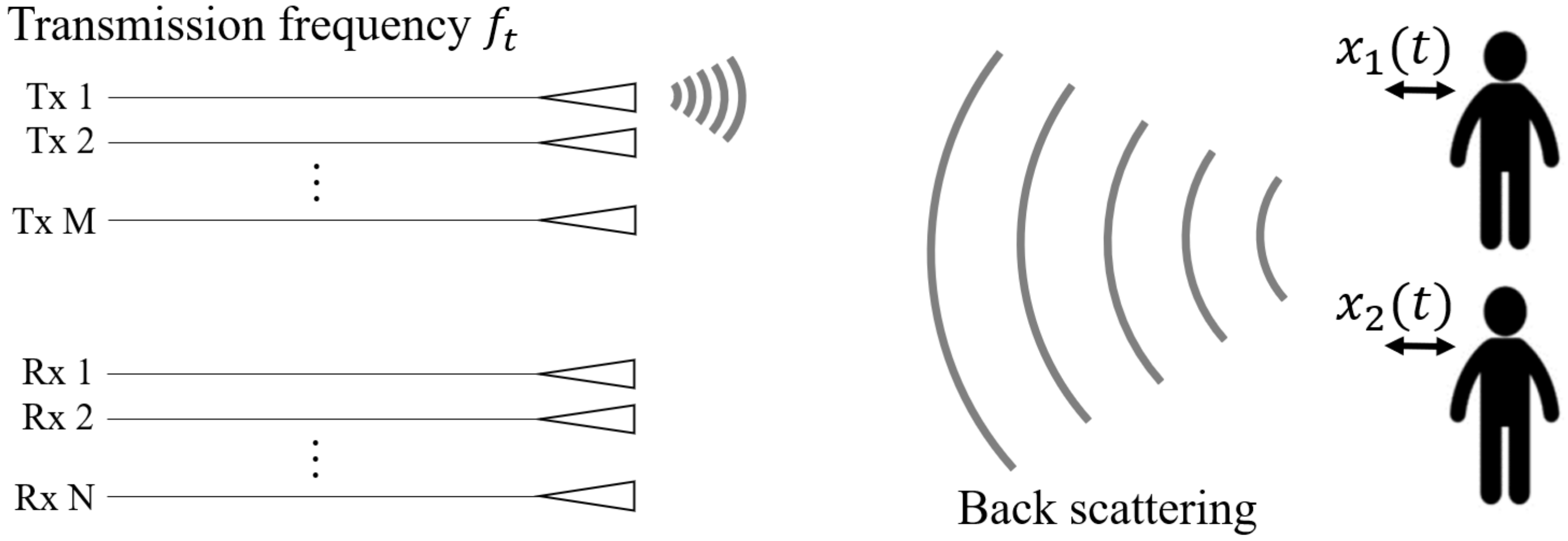}
  \caption{Conceptual illustration of Doppler radar.}
  \label{fig:doppler_radar}
\end{figure}

Fig.~\ref{fig:doppler_radar} is a conceptual illustration showing a measurement scene in an environment without obstacles.
First, a microwave is radiated from Transmitting Antenna Tx1 propagating to a target area. After backscattered on the body surface, it is received by receiving antennas Rx1,\,Rx2, and so on.
Next, microwave is radiated from Transmitting Antenna Tx2, and received in the same way.
In such a manner, the measurement proceeds with the transmitting antennas changed in turn.
The CW Doppler radar to be proposed here is a system that detects chest movements as  the displacement due to respiration and heartbeat.

The phase $\Phi(t)$ of the received microwave of frequency $f_t$ is expressed in terms of the target displacement $x(t)$ as
\begin{equation}
  \label{eq:phi}
  \Phi (t) = 2\pi f_t t + \frac{4\pi x(t)}{\lambda} + \phi_0
\end{equation}
where $\lambda$ and $\phi_0$ represent the wavelength and the phase offset, respectively.

The complex amplitude is calculated from the in-phase component $I(t)$ and the quadrature component $Q(t)$ obtained by an orthogonal detection circuit.
Ignoring the phase offset, we can write a received signal as
\begin{eqnarray}
	& & A(t) \cos\left[\frac{4\pi x(t)}{\lambda} \right] 
	    + j A(t) \sin\left[\frac{4\pi x(t)}{\lambda} \right] 
	\equiv I(t) + j Q(t) 
	\equiv  A(t) \exp \left[j\frac{4\pi x(t)}{\lambda}\right]
	\label{eq:AexpPhi}
\end{eqnarray}
where $A(t)$ is signal amplitude and $j$ is the imaginary unit.
Typically, human respiration causes displacement of about $1~\mathrm{cm}$ to be detected as the phase change.

\subsection{ICA}
\label{subsec:ica}

ICA estimates unmixed original signals only from mixed signal information.
Suppose that $p$ receiving antennas get mixed signals $\bm{x}(t)\in \mathbb{C}^{p\times 1}$ after a mixing matrix $\mathbf{A}\in \mathbb{C}^{p\times p}$ instantaneously mixes original complex signals $\bm{s}(t)\in \mathbb{C}^{p\times 1}$ generated at $p$ signal sources as
\begin{equation}
  \label{eq:x}
  \bm{x}(t) = \mathbf{A}\bm{s}(t).
\end{equation}

It is desired to find a separation matrix $\mathbf{B}\in \mathbb{C}^{p\times p}$ that transforms mixed signals $\bm{x}(t)$ into statistically independent signals $\bm{y}(t)\equiv [y_1(t)\cdots y_p(t)] ^{\mathrm{T}}$, where $[\cdot ]^{\mathrm{T}}$ denotes transposition, as
\begin{equation}
  \label{eq:y}
  \bm{y}(t) = \mathbf{B}\bm{x}(t).
\end{equation}
Each signal of $\bm{y}(t)$ corresponds to one of the original signals $\bm{s}(t)\equiv [s_1(t)\cdots s_p(t)] ^{\mathrm{T}}$.
The separation matrix $\mathbf{B}$ is a solution of ICA.

Basically, the ICA algorithm consists of two parts, namely, whitening and independence maximization.
Whitening is a transformation which makes the data uncorrelated with one another, its mean be 0, and the variance be 1.
This process is closely related to principal component analysis (PCA).
First, a linear transformation for a sample vector $\bm{x}(t)$ at $t$ is considered as
\begin{equation}
  \label{eq:y_of_pca}
  z_i(t) = \bm{v}_i^\mathrm{T} \bm{x}(t),\quad \|\bm{v}_i\| = 1.
\end{equation}
This transformation matrix $\mathbf{V}=[\bm{v}_i]$ that solves the PCA requires two conditions. The first is to select $\bm{v}_1$ that maximizes the variance of the first principal component $z_1(t)=\bm{v}_1^T\bm{x}(t)$. The second is to select $\bm{v}_i$ that $i$-th principal component $z_i(t)$ is uncorrelated with $z_k(t)\quad (k\neq i)$ and the variance is maximized.
The solution to this problem is given by the eigenvalue decomposition of the covariance matrix $\mathbf{C}_{\bm{x}} = \mathrm{E}\{ \bm{x}(t)\bm{x}(t)^\mathrm{T}\}$ of $\bm{x}(t)$ where $E\{\cdot\}$ denotes the average in time.
PCA transformation is given for the eigenvalues $d_1,\, d_2\, \cdots \, d_n \quad (d_1\geq d_2\geq \cdots \geq d_n)$ of $\mathbf{C}_{\bm{x}}$ and the corresponding eigenvectors $\bm{e}_1,\, \bm{e}_2,\, \cdots \bm{e}_n\quad (\| \bm{e}_i\| = 1)$ as
\begin{equation}
  \label{eq:pcatrans}
  \bm{v}_i = \bm{e}_i.
\end{equation}

Whitening is essentially considered as a combination of uncorrelation process achieved by PCA and normalization of the variance to unity.
It is needed to find the whitening matrix $\mathbf{V}$ that whitens $\bm{x}$ into $\bm{z}$ as
\begin{equation}
  \label{eq:z}
  \bm{z} = \mathbf{V}\bm{x}.
\end{equation}
The matrix $\mathbf{V}$ is given as
\begin{equation}
  \label{eq:v_example}
  \mathbf{V} = \mathbf{D}^{-\frac{1}{2}}\mathbf{E}^{\mathrm{T}}
\end{equation}
where $\mathbf{D^{-\frac{1}{2}}} \equiv \mathrm{diag}(d_1^{-\frac{1}{2}},\,\cdots\, ,d_n^{-\frac{1}{2}})$ for the diagonal matrix $\mathbf{D} = \mathrm{diag}(d_1,\,\cdots\, ,d_n)$ and an eigenvector matrix $\mathbf{E} = [e_1,\,\cdots\, ,e_n]$ of $\mathbf{C}_{\bm{x}}$.

Next, the independence maximization transforms the uncorrelated data to independent ones.
Note that uncorrelatedness mentioned above does not necessarily mean independence.
A rotation matrix $\mathbf{W}$ which transforms the whitened and normalized data to independent ones is required.
Here, nonlinear uncorrelatedness can be used as a measure of independence.
If arbitrary variables $y_1$ and $y_2$ are independent, the following theorem holds for arbitrary two functions $h_1$ and $h_2$:
\begin{equation}
  \label{eq:independence}
  \mathrm{E}\{ h_1(y_1)h_2(y_2)\} = \mathrm{E}\{ h_1(y_1)\}\mathrm{E}\{ h_2(y_2)\}.
\end{equation}
From the theorem, if we choose the nonlinear functions $h_1$ and $h_2$ properly, the measure of independence can be determined.
In other words, it is estimated that $y_1$ and $y_2$ are independent if $h_1(y_1)$ and $h_2(y_2)$ are uncorrelated.
In actual algorithms, kurtosis, hyperbolic function $(\mathrm{tanh})$, or another polynomial is often used.

\subsection{Online CF-ICA}
\label{subsec:onlinecfica}

\begin{figure}[H]
  \centering
  \includegraphics[width=0.8\linewidth]{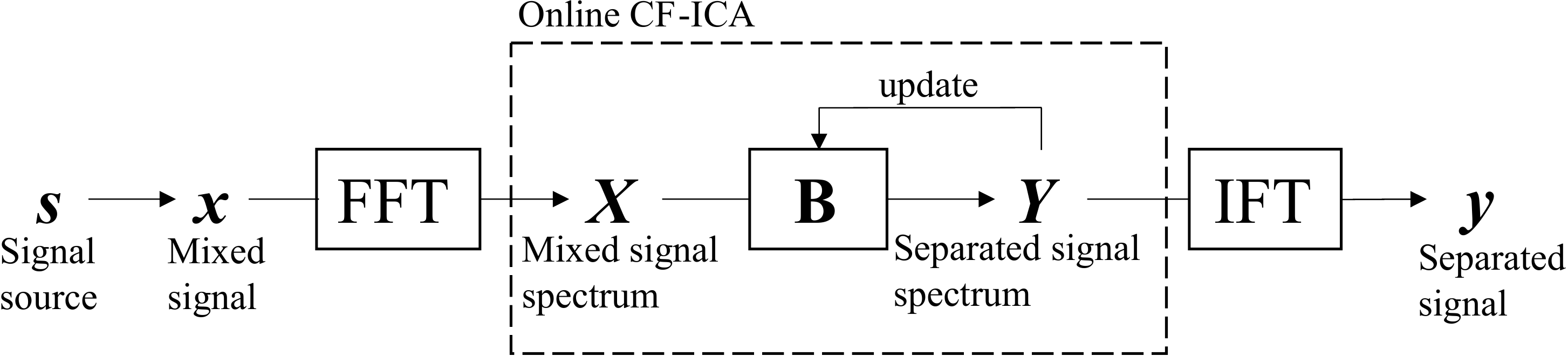}
  \caption{Flow of online CF-ICA.}
  \label{fig:flow}
\end{figure}

Online CF-ICA is an online processing ICA that handles complex signals in the frequency domain \cite{igarss}.
The flow of the process is shown in Fig.~\ref{fig:flow}.
The algorithm described below assumes that the ICA model has no frequency dependence because the relative bandwidth of the microwave CW radar signal is very small in the case of respiration and/or heartbeat measurement.
In general, online ICA learns the separation matrix $\mathbf{B}$ for time series signals that are fed one after another.
Online CF-ICA's online algorithm is based on so-called equivariant adaptive separation via independence (EASI) \cite{cardoso1996equivariant}.
EASI is a method to simultaneously execute the two ICA processes described in Section \ref{subsec:ica}, whitening and independence, in a single updating formula.
First, if the whitening matrix is $\mathbf{V}$, it generates white signals $\bm{z}$ from $\bm{x}$ as
\begin{equation}
  \label{eq:z_easi}
  \bm{z} = \mathbf{V}\bm{x}.
\end{equation}
The updating  fraction for the whitening matrix $\mathbf{V}$ is given as
\begin{equation}
  \label{eq:v_update}
  \Delta\mathbf{V} = \mu_v(\mathbf{I}-\bm{z}\bm{z}^{\mathrm{T}})
\end{equation}
where $\mu_v$ is a learning rate.
Since it is now $\bm{z} = \mathbf{VA}\bm{s}$, the matrix $\mathbf{VA}$ is an orthogonal matrix that gives rotation to realize whitening.

Next, the independence-maximization matrix $\mathbf{W}$ is introduced.
This gives a separation $\bm{y} = \mathbf{W}\bm{z}$ for the whitened signal $\bm{z}$.
At this time, the inverse matrix of $\mathbf{VA}$ can be one of $\mathbf{W}$s.
Thus, $\mathbf{W}$ is also an orthogonal matrix.
To incorporate the measure of independence, the updating fraction is represented as $\Delta\mathbf{W} = \mathbf{D}\mathbf{W}$ by using $\mathbf{D} \equiv -\mu_w g(\bm{y})\bm{y}^{\mathrm{T}}$ including a nonlinear function $g(\cdot)$.
The condition for maintaining the orthogonality of $\mathbf{W}$ in each iteration is written as
\begin{eqnarray}
	(\mathbf{W} + \mathbf{DW})(\mathbf{W} + \mathbf{DW})^{\mathrm{T}} &=& \mathbf{I} + \mathbf{D} + \mathbf{D}^{\mathrm{T}} + \mathbf{D}\mathbf{D}^{\mathrm{T}}
	\nonumber \\
	&=& \mathbf{I}.
	\label{eq:w_update_orthogonality}
\end{eqnarray}
If $\mathbf{D}$ is minute, $\mathbf{D} = -\mathbf{D}^{\mathrm{T}}$ by the first-order approximation.
Coefficient matrix $\mathbf{D}$ is redefined to satisfy this approximation. Then we derive the updating function for $\mathbf{W}$ as
\begin{equation}
  \label{eq:D}
  \mathbf{D} \equiv -\mu_w[g(\bm{y})\bm{y}^{\mathrm{T}} - \bm{y}g(\bm{y})^{\mathrm{T}}],
\end{equation}
\begin{equation}
  \label{eq:delta_W}
  \Delta\mathbf{W} = \mathbf{DW} = -\mu_w[g(\bm{y})\bm{y}^{\mathrm{T}} - \bm{y}g(\bm{y})^{\mathrm{T}}]\mathbf{W}.
\end{equation}
Since the process $\bm{y} = \mathbf{W}\bm{z} = \mathbf{WV}\bm{x}$ performs the above whitening and independence maximization in series, the separation matrix is represented as $\mathbf{B} = \mathbf{WV}$.
Combining the updating fractions (\ref{eq:v_update}) and (\ref{eq:delta_W}) derives the learning rule of the separation matrix $\mathbf{B}$ as
\begin{eqnarray}
	\mathbf{B} &\longleftarrow& \mathbf{B} + \Delta\mathbf{B},
	\\
	\Delta\mathbf{B}&=&\Delta\mathbf{WV} + \mathbf{W}\Delta\mathbf{V}
	\nonumber \\
	&=& -\mu_v[g(\bm{y})\bm{y}^{\mathrm{T}} - \bm{y}g(\bm{y})^{\mathrm{T}}]\mathbf{WV}
	+ \mu_w[\mathbf{WV} - \mathbf{W}\bm{z}\bm{z}^{\mathrm{T}}\mathbf{W}^{\mathrm{T}}\mathbf{WV}]
	\nonumber \\
	&=&-\mu[\bm{y}\bm{y}^{\mathrm{T}} - \mathbf{I} + g(\bm{y})\bm{y}^{\mathrm{T}} - \bm{y}g(\bm{y})^{\mathrm{T}}]\mathbf{B}
	\label{eq:B_update}
\end{eqnarray}
where the learning rates in (\ref{eq:v_update}) and (\ref{eq:delta_W}) are $\mu_v = \mu_w = \mu$.
The complex EASI is obtained by changing $[\cdot]^{\mathrm{T}}$ to transpose conjugate $[\cdot]^{\mathrm{H}}$ as
\begin{equation}
  \label{eq:comp_deltaB}
  \Delta\mathbf{B} = -\mu[\bm{y}\bm{y}^{\mathrm{H}} - \mathbf{I} + g(\bm{y})\bm{y}^{\mathrm{H}} - \bm{y}g(\bm{y})^{\mathrm{H}}]\mathbf{B}.
\end{equation}
The online CF-ICA is an extension of this operation to the frequency domain.
The frequency-domain processing has an advantage of easiness to narrow down the frequency band of the signals used in the learning for noise elimination.

Online CF-ICA uses short-time Fourier transform (STFT) to convert time domain signals into frequency domain.
Since respiration and heartbeat are the measurement targets, the frequency band we are interested in is from $f_{\mathrm{min}} = 0.17~\mathrm{Hz}$ to $f_{\mathrm{max}} = 2.0~\mathrm{Hz}$.
The time-domain signals $\bm{x}(t)\equiv[x_i(t)]$ and $\bm{y}(t)\equiv[y_i(t)]$ are transformed by STFT as
\begin{equation}
  \label{eq:X_stft}
  \bm{X}(\omega, t_d) = [X_i(\omega, t_d)] = \left[ \sum_{\tau=0}^{L_{\mathrm{STFT}}-1} x_i(\tau + t_dS)e^{-j\omega\tau} \right],
\end{equation}
\begin{equation}
  \label{eq:Y_stft}
  \bm{Y}(\omega, t_d) = [Y_i(\omega, t_d)] = \left[ \sum_{\tau=0}^{L_{\mathrm{STFT}}-1} y_i(\tau + t_dS)e^{-j\omega\tau} \right]
\end{equation}
where $L_{\mathrm{STFT}}$ is the length of the Fourier window, $S$ is the moving step of the Fourier window, and $t_d$ is the discrete time.
In the instantaneous mixing case, the linearity of the Fourier transform assures the applicability of the time-domain ICA model expressed by (\ref{eq:y}) also in the frequency domain.
Thus, the following separation learning is possible for angular frequency $\omega$ ($2\pi f_{\mathrm{min}} < \omega <2\pi f_{\mathrm{max}}$) as
\begin{equation}
  \label{eq:Y_freq}
  \bm{Y}(\omega, t_d) = \mathbf{B}(t_d)\bm{X}(\omega, t_d),
\end{equation}
\begin{equation}
  \label{eq:comp_deltaB_freq}
  \Delta\mathbf{B} = -\mu[\bm{Y}\bm{Y}^{\mathrm{H}} - \mathbf{I} + g(\bm{Y})\bm{Y}^{\mathrm{H}} - \bm{Y}g(\bm{Y})^{\mathrm{H}}]\mathbf{B}.
\end{equation}
As mentioned before, $\mathbf{B}$ does not depend on frequency.

In the update of $\mathbf{B}$ by using $\Delta\mathbf{B}$ in (\ref{eq:comp_deltaB_freq}), we apply an appropriate scaling of $\mathbf{B}$ and/or $\bm{Y}$ practically because of the following two reasons.
The first is that the component values of $\mathbf{B}$ can overflow in a computer.
The second is that it is important to effectively use the nonlinear region of the nonlinear function $g(\cdot)$.
The following process achieves the scaling.

We focus on $g(\bf{Y})$$\,= [g(Y_i)]$.
We want each component $Y_i$ of $\bm{Y} = \mathbf{B}\bm{X}$ be in a desired range of $g(\cdot)$ by scaling $\bf{B}$.
We consider an algorithm for each short data window instead of for each sample.
We calculate the root mean square (RMS) of the signal in a window as
\begin{equation}
  \label{eq:RMS}
  \mathrm{RMS}_i = \sqrt{\frac{1}{L_{\mathrm{STFT}}}\sum_{\tau=0}^{L_{\mathrm{STFT}}-1}|Y_i(\tau)|^2}
\end{equation}
to scale $\mathbf{B}$ as
\begin{equation}
  \label{eq:B_scaling}
  \mathbf{B} \longleftarrow \mathrm{diag}[\mathrm{RMS}_i^{-1}]\mathbf{B}.
\end{equation}
For the learning, $\bm{Y} = \mathbf{B}\bm{X}$ is calculated after scaling $\mathbf{B}$.
As a result, the RMS of $Y_i(t)$ becomes unity in this window.
With this $\bm{Y}$, the training within the window is performed according to (\ref{eq:comp_deltaB_freq}).
Repeating the above scaling makes the nonlinearity of $g(\cdot)$ always effective regardless of the magnitude of the received signals.

\section{Proposal of higher-order tensor independent component analysis (HOT-ICA)}
\label{sec:onlinehotica}

\begin{figure}[H]
  \centering
  \includegraphics[width=0.6\linewidth]{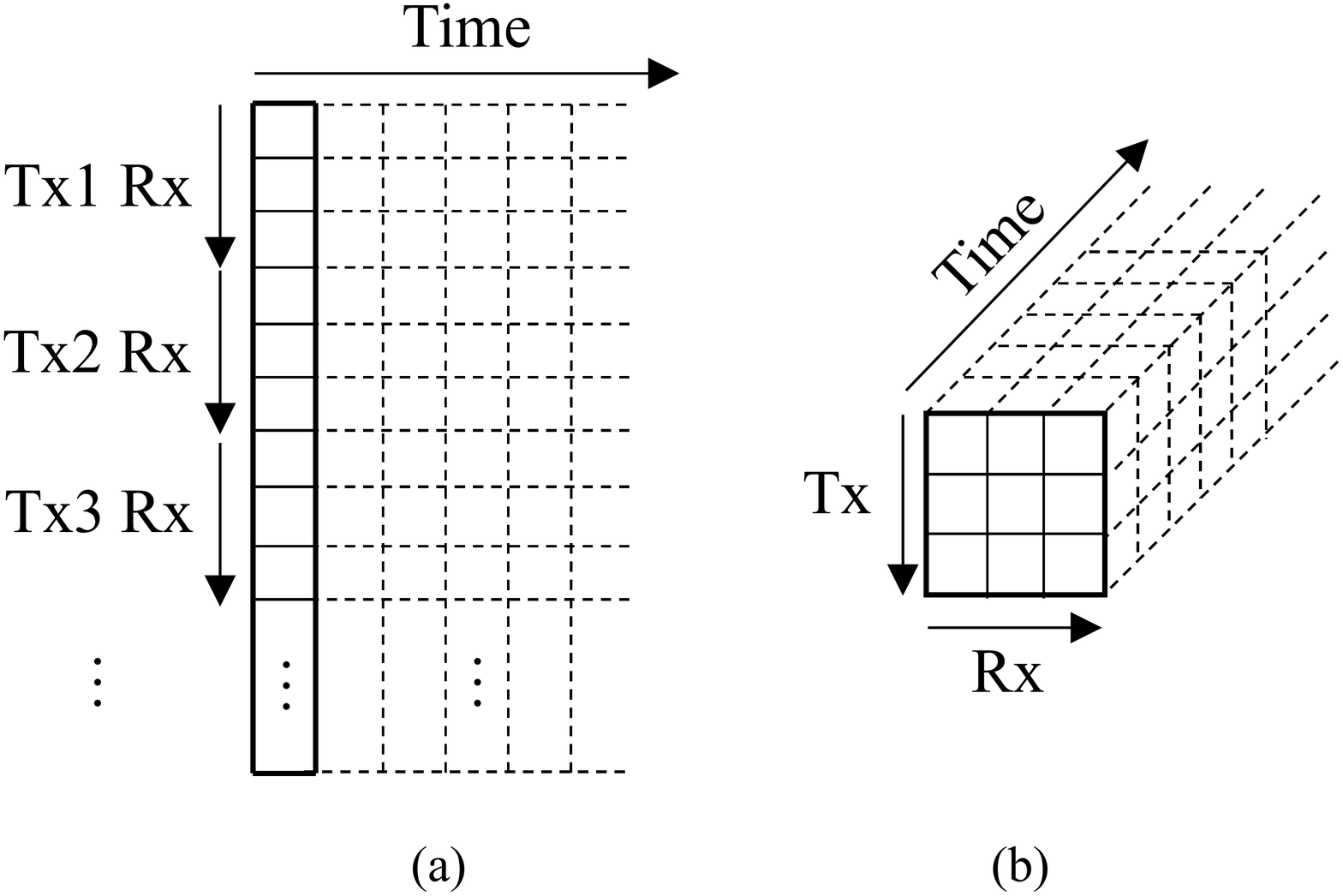}
  \caption{Structures of received data tensors for (a) conventional ICA and (b) HOT-ICA in their processing.}
  \label{fig:datashape}
\end{figure}

As mentioned in Section~\ref{s:introduction}, we propose HOT-ICA as an ICA method completely different from the tensor decomposition.
In the field of  heartbeat/respiration remote sensing, HOT-ICA is a method effective in particular in MIMO systems.

Fig.~\ref{fig:datashape} compares the data forms used in (a) conventional ICA and (b) HOT-ICA, respectively. In Fig.~\ref{fig:datashape} (a), the data is represented in a vector to be used with matricization. This reformation nullifies the data categories indicated by the tensor axes. In contrast, in Fig.~\ref{fig:datashape} (b), the Tx and Rx axes remain effective so that we can utilize the information on which transmitting/receiving antennas generated the data.

Let the number of transmitting antennas be $p_t$, the number of receiving antennas be $p_r$, and the obtained mixed signals be $\mathbf{x}(t)\equiv x(t)^{(\gamma,\delta)}\in \mathbb{C}^{p_t\times p_r}$.
There, we find an equation that transforms the mixed signals into statistically independent signals $\mathbf{y}(t)\equiv y(t)^{(\alpha,\beta)}\in \mathbb{C}^{p_t\times p_r}$.
By referring to (\ref{eq:y}), we adopt a fourth order tensor $\underline{\underline{\mathbf{B}}}\equiv B^{(\alpha,\beta,\gamma,\delta)}\in \mathbb{C}^{p_t\times p_r\times p_t\times p_r}$, with $\underline{\underline{\mathbf{\ \cdot\ }}}$ denote a fourth order tensor, for the transformation expressed as
\begin{equation}
  \label{eq:4ordery}
  y(t)^{(\alpha,\beta)} = \sum_{\gamma = 1}^{p_t} \sum_{\delta = 1}^{p_r} B^{(\alpha,\beta,\gamma,\delta)} x(t)^{(\gamma,\delta)}.
\end{equation}
The HOT-ICA model (\ref{eq:4ordery}) in the time domain is also applicable in the frequency domain by using STFT, just like the online CF-ICA, as
\begin{equation}
  \label{eq:4ordery_freq}
 Y(\omega, t_d)^{(\alpha,\beta)} = \sum_{\gamma = 1}^{p_t} \sum_{\delta = 1}^{p_r} B(t_d)^{(\alpha,\beta,\gamma,\delta)} X(\omega, t_d)^{(\gamma,\delta)}.
\end{equation}

We extend the updating formula (\ref{eq:comp_deltaB_freq}) to HOT-ICA calculation to construct a new updating procedure as
\begin{equation}
  \label{eq:comp_deltaB_freq_hotica}
  \Delta B^{(\alpha,\beta,\gamma,\delta)} = \sum_{\varepsilon = 1}^{p_t}\sum_{\zeta = 1}^{p_r} W^{(\alpha,\beta,\varepsilon,\zeta)}\,B^{(\varepsilon,\zeta,\gamma,\delta)}
\end{equation}
where we define a learning weight tensor $\underline{\underline{\mathbf{W}}}\equiv W^{(\alpha,\beta,\gamma,\delta)}\in \mathbb{C}^{p_t\times p_r \times p_t\times p_r}$ as
\begin{eqnarray}
  \label{eq:comp_W_freq_hotica}
  W^{(\alpha,\beta,\gamma,\delta)}
   &=&-\mu \left[ Y^{(\alpha,\beta)} \bar{Y}^{(\gamma,\delta)} +g(Y^{(\alpha,\beta)}) \bar{Y}^{(\gamma,\delta)} 
   +Y^{(\alpha,\beta)} \bar{g(Y)}^{(\gamma,\delta)} - I^{(\alpha,\beta,\gamma,\delta)} \right]
\end{eqnarray}
where we also define $\underline{\underline{\mathbf{I}}}\in \mathbb{C}^{p_t\times p_r \times p_t\times p_r}$ as
\begin{equation}
  \label{eq:4order_idn}
  I^{(\alpha,\beta,\gamma,\delta)}
  = \left \{
 \begin{array}{l}
 1 \quad(\alpha = \gamma \cap \beta = \delta), \\
 0 \quad(\alpha \neq \gamma \cup \beta \neq \delta).
 \end{array}
 \right. 
\end{equation}


When we construct a MIMO system, it is inevitable that respective antennas have various conditions and/or situations different from one another depending on the environment. For example, an antenna with an amplifier may be relatively noisy or defective.
In such a case, we should improve the robustness of the overall learning process.
This can be achieved by reducing the learning weight associated with the defective antenna.
The HOT-ICA can do it as follows. We break down the updating formula (\ref{eq:comp_deltaB_freq_hotica}).
By assuming that the updating tensor components related to the transmitting antennas Tx1,\,Tx2,\,...\,,\,Tx$M$ are $\Delta\underline{\underline{\mathbf{B}}}_{\mathrm{Tx1}},\,\Delta\underline{\underline{\mathbf{B}}}_{\mathrm{Tx2}},\,...\,,\,\Delta\underline{\underline{\mathbf{B}}}_{\mathrm{Tx}M}$ and those related to the receiving antennas Rx1,\,Rx2,\,...\,,\,Rx$N$ are $\Delta\underline{\underline{\mathbf{B}}}_{\mathrm{Rx1}},\,\Delta\underline{\underline{\mathbf{B}}}_{\mathrm{Rx2}},\,...\,,\,\Delta\underline{\underline{\mathbf{B}}}_{\mathrm{Rx}N}$, we can express $\Delta\underline{\underline{\mathbf{B}}}$ as
\begin{eqnarray}
  \label{eq:deltaB_decomp}
  \Delta\underline{\underline{\mathbf{B}}}
  = 
  \frac{1}{2}\left( \Delta\underline{\underline{\mathbf{B}}}_{\mathrm{Tx1}} + \Delta\underline{\underline{\mathbf{B}}}_{\mathrm{Tx2}}
  + \cdots + \Delta\underline{\underline{\mathbf{B}}}_{\mathrm{Tx}M}
  + \Delta\underline{\underline{\mathbf{B}}}_{\mathrm{Rx1}} + \Delta\underline{\underline{\mathbf{B}}}_{\mathrm{Rx2}}
  + \cdots + \Delta\underline{\underline{\mathbf{B}}}_{\mathrm{Rx}N} \right)
\end{eqnarray}
where $\Delta\underline{\underline{\mathbf{B}}}_{\mathrm{Tx}m}\quad(1\leq m\leq M)$ and $\Delta\underline{\underline{\mathbf{B}}}_{\mathrm{Rx}n}\quad(1\leq n\leq N)$ are defined as
\begin{equation}
  \label{eq:comp_deltaB_Txm}
  \Delta B_{\mathrm{Tx}m}^{(\alpha,\beta,\gamma,\delta)} = \sum_{\varepsilon = 1}^{p_t}\sum_{\zeta = 1}^{p_r}W_{\mathrm{Tx}m}^{(\alpha,\beta,\varepsilon,\zeta)}\,B^{(\varepsilon,\zeta,\gamma,\delta)},
\end{equation}
\begin{equation}
  \label{eq:comp_deltaB_Rxn}
  \Delta B_{\mathrm{Rx}n}^{(\alpha,\beta,\gamma,\delta)} = \sum_{\varepsilon = 1}^{p_t}\sum_{\zeta = 1}^{p_r}W_{\mathrm{Rx}n}^{(\alpha,\beta,\varepsilon,\zeta)}\,B^{(\varepsilon,\zeta,\gamma,\delta)}.
\end{equation}
This is possible in HOT-ICA, which keeps tensor axes meaningful.
The learning weight tensors $\underline{\underline{\mathbf{W}}}_{\mathrm{Tx}m}$ and $\underline{\underline{\mathbf{W}}}_{\mathrm{Rx}n}$ are represented as
\begin{equation}
  \label{eq:comp_W_Txm}
  W_{\mathrm{Tx}m}^{(\alpha,\beta,\gamma,\delta)}
  = \left \{
 \begin{array}{l}
 W^{(\alpha,\beta,\gamma,\delta)} \quad(\gamma = m), \\
 0 \quad\quad\quad\quad\ (\gamma \neq m),
 \end{array}
 \right. 
\end{equation}
\begin{equation}
  \label{eq:comp_W_Rxn}
  W_{\mathrm{Rx}n}^{(\alpha,\beta,\gamma,\delta)}
  = \left \{
 \begin{array}{l}
 W^{(\alpha,\beta,\gamma,\delta)} \quad(\delta = n), \\
 0 \quad\quad\quad\quad\ (\delta \neq n).
 \end{array}
 \right. 
\end{equation}
For example, a coefficient $\eta_{\mathrm{Rx}1}\quad(0\leqq\eta_{\mathrm{Rx}1}<1)$ can individually reduce the weight $\underline{\underline{\mathbf{W}}}_{\mathrm{Rx}1}$ of the learning weight tensor for a defective antenna Rx1 to obtain a new tensor
\begin{equation}
  \label{eq:reduc_W_Rx1}
  \underline{\underline{\mathbf{W}}}_{\mathrm{Rx}1}^{\prime} = \eta_{\mathrm{Rx}1} \underline{\underline{\mathbf{W}}}_{\mathrm{Rx}1}
\end{equation}
to control the sensitivity partially for a higher robustness in the learning.

In this way, HOT-ICA can manipulate the learning sensitivity for some of the components associated with respective antenna situations.
This is very effective for measurements employing the MIMO configuration.
Conventional methods such as online CF-ICA (see Section~\ref{subsec:onlinecfica}) cannot perform this manipulation.

Note that tensor calculation of HOT-ICA is different from that of MICA based on the Tucker decomposition, which requires matricization.
HOT-ICA keeps the data tensor structure without nullifying the categorization.
Hence, HOT-ICA is capable of adaptive signal-source separation every time the receiving antennas acquire signals even including possible changes in the measurement environment, resulting in an enhanced robustness.
Note also that this proposal is extendable to a processing for $n$-th order mixed and unmixed signals by use of a $2n$-th order separation tensor.

\section{Experimental setup}
\label{s:numerical}

\begin{figure}[H]
  \centering
  \includegraphics[width=0.8\linewidth]{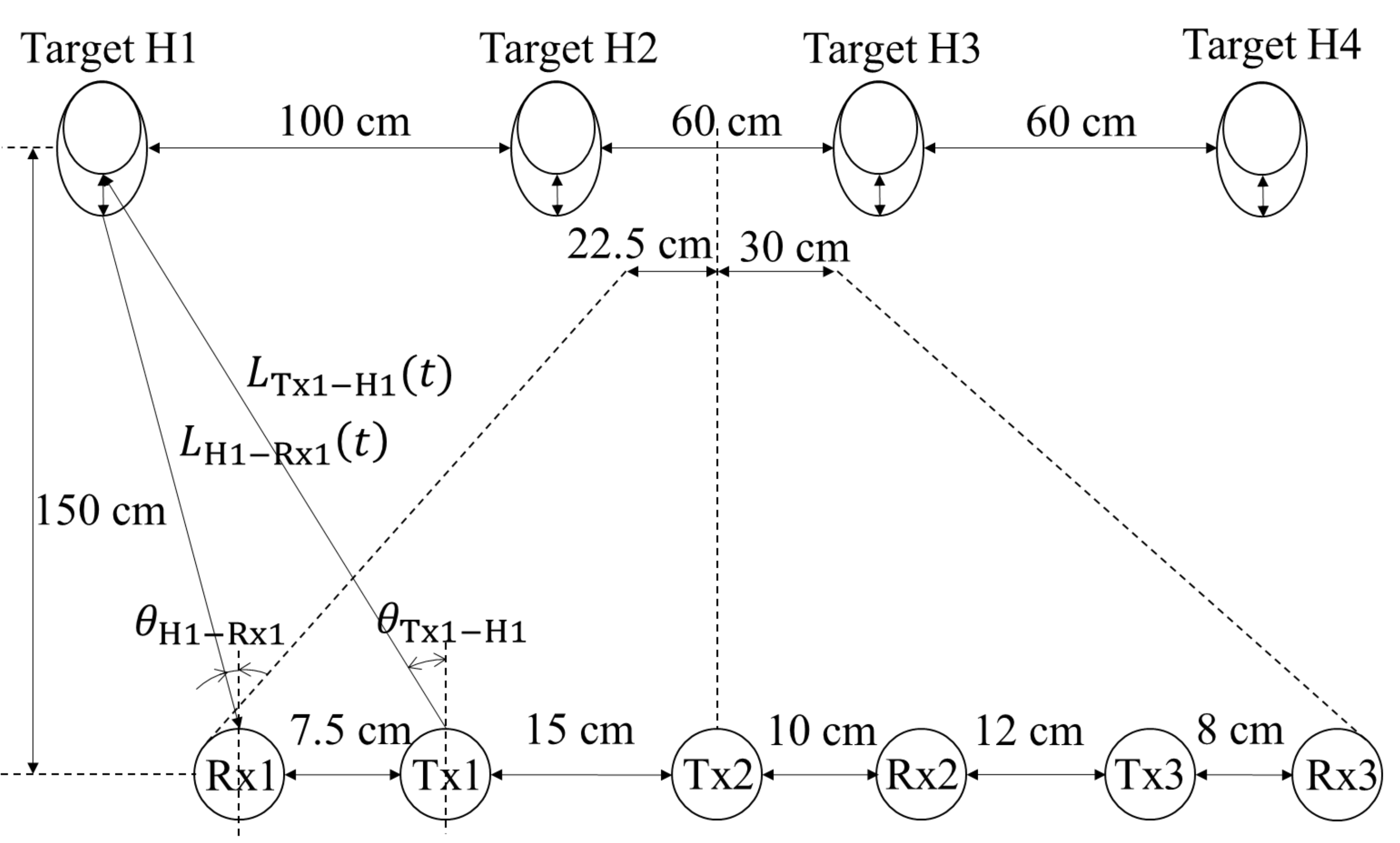}
  \caption{Placement of transmitting and receiving antennas, Tx$m$ and Rx$n$, and target humans H$k$.}
  \label{fig:placement}
\end{figure}

We conduct a numerical-physical experiment by assuming a CW MIMO Doppler radar front-end 
for signal-source separation based on the HOT-ICA.
Fig.~\ref{fig:placement} shows the placement of antennas and targets.
The numbers of transmitting antennas Tx and receiving antennas Rx are $p_t=p_r=3$, and the number of targets (humans: H) is 4.
At the humans, the chest moves periodically due to respiration and heartbeat, and the Doppler radar detects the body displacement. Then, complex-valued signals are finally obtained.
The signals consist of not only the signals originating from the targets but also various noise.

\begin{table}[H]
 \centering
 \caption{Parameters related to respiration and heartbeat of each target}
 \label{tb:param_of_target}
 \begin{tabular}{l|l|l|l|l|l}
  \hline
  \multicolumn{2}{l|}{} & Target H1 & Target H2 & Target H3 & Target H4 
  \\ \hline \hline
  \multirow{2}{*}{Respiration}  & Amplitude & $a_{r_1}=0.5\times 10^{-3}$ & $a_{r_2}=0.5\times 10^{-3}$ & $a_{r_3}=0.5\times 10^{-3}$ & $a_{r_4}=0.5\times10^{-3}$
  \\ \cline{2-6} 
    & Frequency & $f_{r_1}=0.40$\,Hz & $f_{r_2}=0.31$\,Hz & $f_{r_3}=0.71$\,Hz & $f_{r_4}=0.53$\,Hz 
  \\ \hline
  \multirow{2}{*}{Heartbeat}  & Amplitude & $a_{h_1}=0.05\times 10^{-3}$ & $a_{h_2}=0.04\times 10^{-3}$ & $a_{h_3}=0.06\times 10^{-3}$ & $a_{h_4}=0.03\times10^{-3}$
  \\ \cline{2-6} 
   & Frequency & $f_{h_1}=1.19$\,Hz & $f_{h_2}=1.10$\,Hz & $f_{h_3}=1.32$\,Hz & $f_{h_4}=1.06$\,Hz 
  \\ \hline
 \end{tabular}
\end{table}

\begin{figure}[H]
  \centering
  \includegraphics[width=0.4\linewidth]{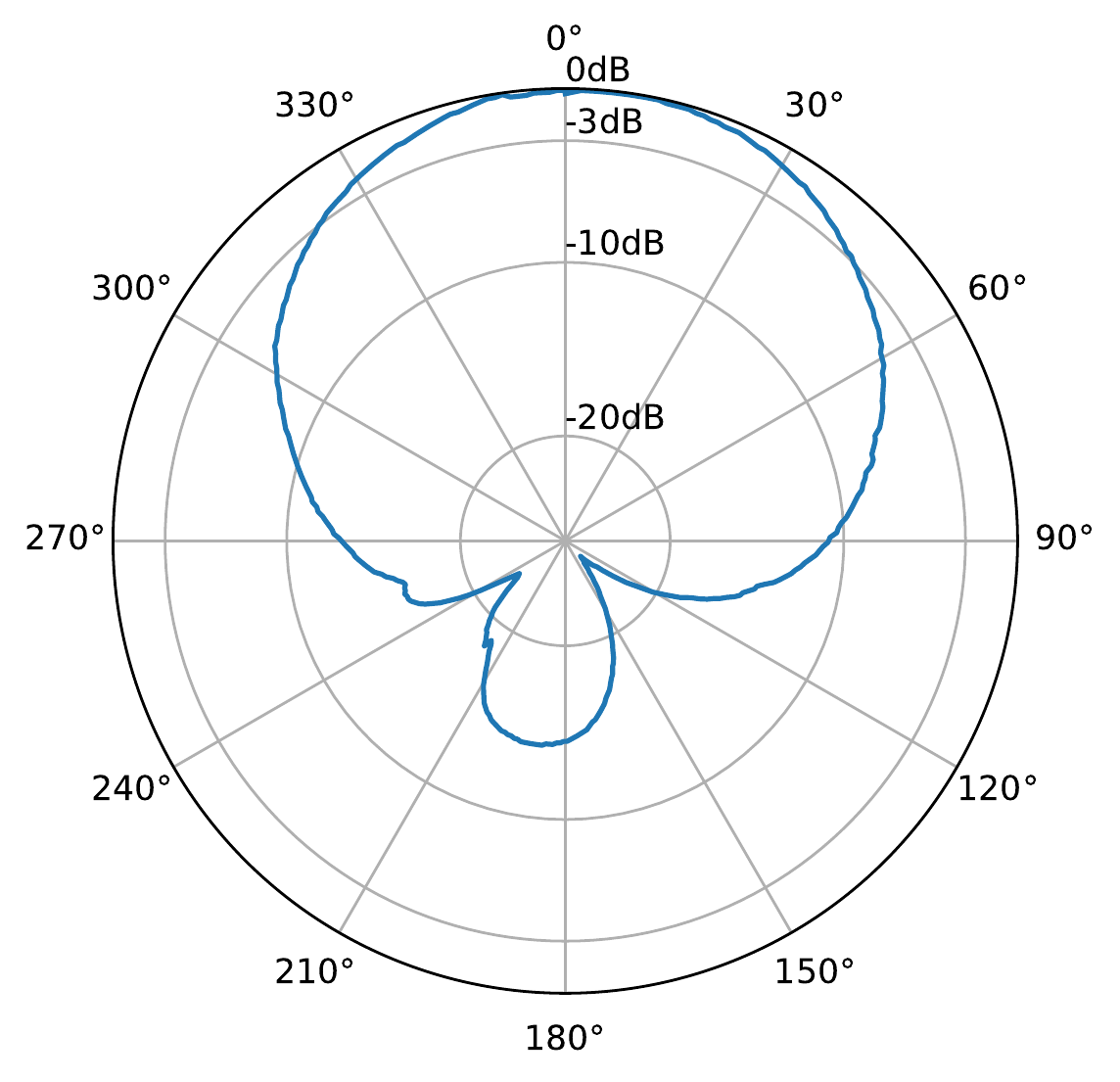}
  \caption{Directivity (gain) of the transmitting and receiving antennas.}
  \label{fig:Directivities}
\end{figure}

In this numerical-physical experiment, $\bar{L}_{\mathrm{Tx}m\textendash\mathrm{H}k}$ denotes the average distance from a transmitting antenna Tx$m\quad(1 \leqq m \leqq 3)$ to a target H$k\quad(1 \leqq k \leqq 4)$, and $\bar{L}_{\mathrm{H}k\textendash\mathrm{Rx}n}$ represents that from a target H$k$ to a receiving antenna Rx$n\quad(1 \leqq n \leqq 3)$.
The direction angle from the transmitting antenna Tx$m$ to the target H$k$ is $\theta_{\mathrm{Tx}m\textendash\mathrm{H}k}$, and the arrival angle from the target H$k$
to the receiving antenna Rx$n$ is $\theta_{\mathrm{H}k\textendash\mathrm{Rx}n}$.
In the following equations, we use $d_{\mathrm{Tx}m}$ and $d_{\mathrm{Rx}n}$ representing the directivities (gains) of the transmitting and receiving antennas shown in Fig.~\ref{fig:Directivities}, and $\sigma$ indicating the scattering coefficient of electromagnetic waves on the human body.

We determine the original signal model $\bm{s}(t)\in \mathbb{C}^{4 \times 1}$ as
\begin{eqnarray}
  \bm{s}(t) = \left[
    \begin{array}{llll}
      s_{\mathrm{H}1} \\
      s_{\mathrm{H}2} \\
      s_{\mathrm{H}3} \\
      s_{\mathrm{H}4}
    \end{array}
   \right]
   = \left[
    \begin{array}{llll}
      \mathrm{exp}(j w_{\mathrm{H}1}) \\
      \mathrm{exp}(j w_{\mathrm{H}2}) \\
      \mathrm{exp}(j w_{\mathrm{H}3}) \\
      \mathrm{exp}(j w_{\mathrm{H}4})
    \end{array}
   \right]
\end{eqnarray}
where $w_{\mathrm{H}1}$, $w_{\mathrm{H}2}$, $w_{\mathrm{H}3}$ and $w_{\mathrm{H}4}$ are
\begin{eqnarray}
  w_{\mathrm{H}1} &=& a_{r_1} \sin(2\pi f_{r_1} t) + a_{h_1} \sin(2 \pi f_{h_1} t),
  \\
  w_{\mathrm{H}2} &=& a_{r_2} \sin(2\pi f_{r_2} t + \pi/6) 
  + a_{h_2} \sin(2 \pi f_{h_2} t + \pi/6),
  \\
  w_{\mathrm{H}3} &=& a_{r_3} \sin(2\pi f_{r_3} t + 3\pi/4) 
  + a_{h_3} \sin(2 \pi f_{h_3} t + 3\pi/4),
  \\
  w_{\mathrm{H}4} &=& a_{r_4} \sin(2\pi f_{r_4} t + \pi) 
  + a_{h_4} \sin(2 \pi f_{h_4} t + \pi)
\end{eqnarray}
expressing respiration and heartbeat signals of the four humans with their amplitudes and frequencies shown in Table~\ref{tb:param_of_target}.

The received signals $\mathbf{E}_{\mathrm{rec}}\in \mathbb{C}^{p_t\times p_r}$ is represented as
\begin{eqnarray}
  \label{eq:recsig_e}
  &&\mathbf{E}_{\mathrm{rec}}(L_{\mathrm{Tx}m\textendash\mathrm{H}k}(t), L_{\mathrm{H}k\textendash\mathrm{Rx}n}(t))^{(m, n)}
  \nonumber\\
  &&\qquad\qquad = \sum_{k=1}^{4} \left[ d_{\mathrm{Tx}m}
  \frac{\mathrm{exp}(j2\pi \frac{L_{\mathrm{Tx}m\textendash\mathrm{H}k}(t)}{\lambda})}{L_{\mathrm{Tx}m\textendash\mathrm{H}k}(t)} 
  \cdot \sigma 
  d_{\mathrm{Rx}n} 
  \frac{\mathrm{exp}(j2\pi \frac{L_{{\mathrm{H}k} \textendash \mathrm{Rx}n}(t)}{\lambda})}{L_{{\mathrm{H}k} \textendash \mathrm{Rx}n}(t)} \right]
\end{eqnarray}
where $L_{\mathrm{Tx}m\textendash\mathrm{H}k}(t)$ and $L_{\mathrm{H}k\textendash\mathrm{Rx}n}(t)$ are written as
\begin{eqnarray}
  L_{\mathrm{Tx}m\textendash\mathrm{H}k}(t) &=& \bar{L}_{\mathrm{Tx}m\textendash\mathrm{H}k} - w_{\mathrm{H}k}\cos\theta_{\mathrm{Tx}m\textendash\mathrm{H}k},
  \\
  L_{\mathrm{H}k\textendash\mathrm{Rx}n}(t) &=& \bar{L}_{\mathrm{H}k\textendash\mathrm{Rx}n} - w_{\mathrm{H}k}\cos\theta_{\mathrm{H}k\textendash\mathrm{Rx}n}.
\end{eqnarray}

In this experiment, the received signal model $\mathbf{E}_{\mathrm{rec}}$ and noise $\mathbf{V}_n\in \mathbb{C}^{p_t\times p_r}$ result in mixed signals $\mathbf{x}(t)=x(t)^{(\gamma,\delta)}$ in (\ref{eq:4ordery}) as
\begin{equation}
  \label{eq:sim_x_time}
  \mathbf{x}(t) = \mathbf{E}_{\mathrm{rec}}(L_{\mathrm{Tx}m\textendash\mathrm{H}k}(t), L_{\mathrm{H}k\textendash\mathrm{Rx}n}(t)) + \mathbf{V}_n,
\end{equation}
\begin{equation}
  \label{eq:V_noise}
  \mathbf{V}_n = \rho \mathbf{R}
\end{equation}
where $\rho$ is a coefficient that determines the magnitude of noise,
$\mathbf{R}\in \mathbb{C}^{p_t\times p_r}$ is a tensor consisting of random value components $R^{(m, n)}$ following the normal distribution with a mean of 0 and a variance of 1.

\begin{table}[H]
	\centering
	\caption{Parameters for HOT-ICA and processing}
	\label{tb:param_sim}
	\begin{tabular}{l|c}
	\hline
		Parameter & Value 
		\\ \hline \hline
		Sampling rate $f_s$ & 11.3 Hz 
		\\ \hline
		STFT Window size $L_{\mathrm{STFT}}$ & 256 
		\\ \hline
		Moving step of windows $S$ & 2 
		\\\hline
		Frequency band $f_{\mathrm{min}}$\textendash$f_{\mathrm{max}}$& 0.17\textendash 2.0 Hz 
		\\ \hline
		Non-linear function $g(s)$ & tanh$|s|$\,exp$(j\,\mathrm{arg}(s))$ 
		\\ \hline
		Learning rate $\mu$ & 0.0050 
		\\ \hline
		Noise coefficient $\rho$ & $0.50 \times 10^{-5}$ 
		\\\hline
		Backscattering coefficient $\sigma$ & 0.068
		\\ \hline
	\end{tabular}
\end{table}

The parameters of HOT-ICA are shown in Table~\ref{tb:param_sim}.
We receive signals for 70~s.
Since the sampling frequency is $f_s = 11.3$~Hz, there are 790 data points.
With the STFT of window size $L_{\mathrm{STFT}} = 256$ and moving step $S = 2$, the total number of STFT outputs is 267.
Then, the discrete time $t_d$ corresponding to every STFT ranges from 0 to $T_d = 267$.

\section{Results and discussion}
\label{s:numericalresult}
\subsection{Signal-source separation by HOT-ICA}

\begin{figure}[H]
  \centering
  \includegraphics[width=0.9\linewidth]{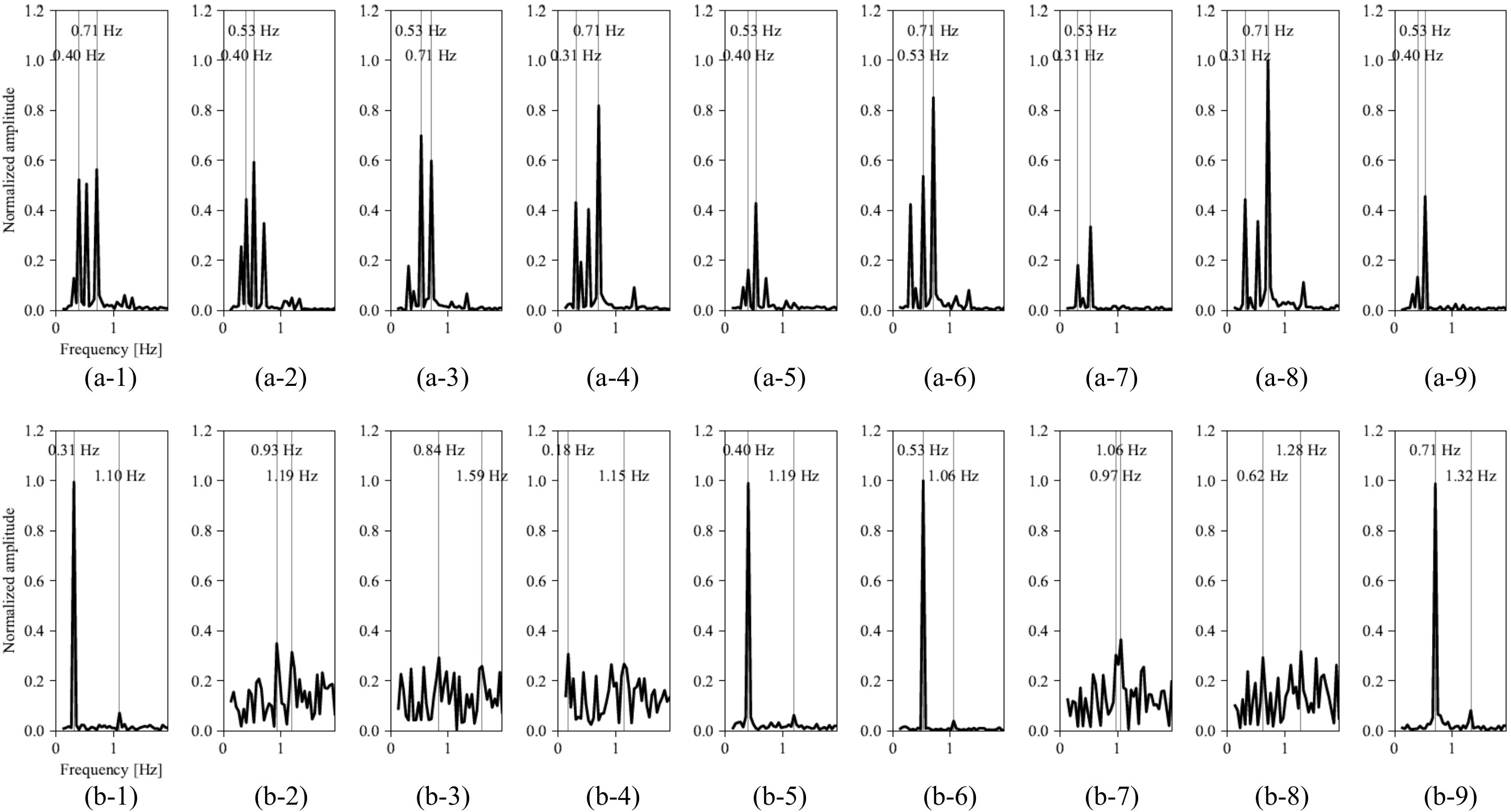}
  \caption{Spectra of (a-$\star$) original mixed signals and (b-$\star$) separated signals obtained by HOT-ICA at the last STFT window for the setting of three transmitting and three receiving antennas.}
  \label{fig:3x3spec}
\end{figure}

We evaluate the performance of the signal-source separation using HOT-ICA.
Fig.~\ref{fig:3x3spec} shows spectra obtained as the learning results at the last time window $(t_d = T_d)$ in the observation for the mixed signals $\mathbf{x}(t)$.
The upper row shows the mixed signals while the lower row presents the separated signals.
The vertical axis represents the signal magnitude normalized in such a way that the maximum signal magnitude in each row becomes unity, and the horizontal axis indicates the frequency.

From left to right, the upper graphs in Fig.~\ref{fig:3x3spec} show the mixed signal spectra $\mathbf{X}(\omega,\,T_d)$ obtained with (Transmitting Antenna Tx1, Receiving Antenna Rx1), (Tx1, Rx2), (Tx1, Rx3), (Tx2, Rx1), (Tx2, Rx2), (Tx2, Rx3), (Tx3, Rx1), (Tx3, Rx2) ), and (Tx3, Rx3).
They present the sum of the original signals scattered at the four targets, the measurement environment, and the noise at the amplifiers.
The spectra in the lower row are the separated signal spectra $\mathbf{Y}(\omega,\,T_d)$ obtained by the HOT-ICA.
In each of Figs.~\ref{fig:3x3spec} (b-1), (b-5), (b-6) or (b-9), we can observe a large primary peak and a small secondary peak in the signal magnitude.
In Table~\ref{tb:param_of_target}, the frequencies of the primary and secondary peaks correspond to those of respiration and heartbeat of respective targets, and we find that the primary peak represents respiration while the secondary indicates heartbeat.
Note that respiration and heartbeat signals of each target appear simultaneously in a single spectrum.
Other spectra in Figs.~\ref{fig:3x3spec} (b-2), (b-3), (b-4), (b-7) and (b-8) present noise only.

\begin{figure}[H]
  \centering
  \includegraphics[width=0.9\linewidth]{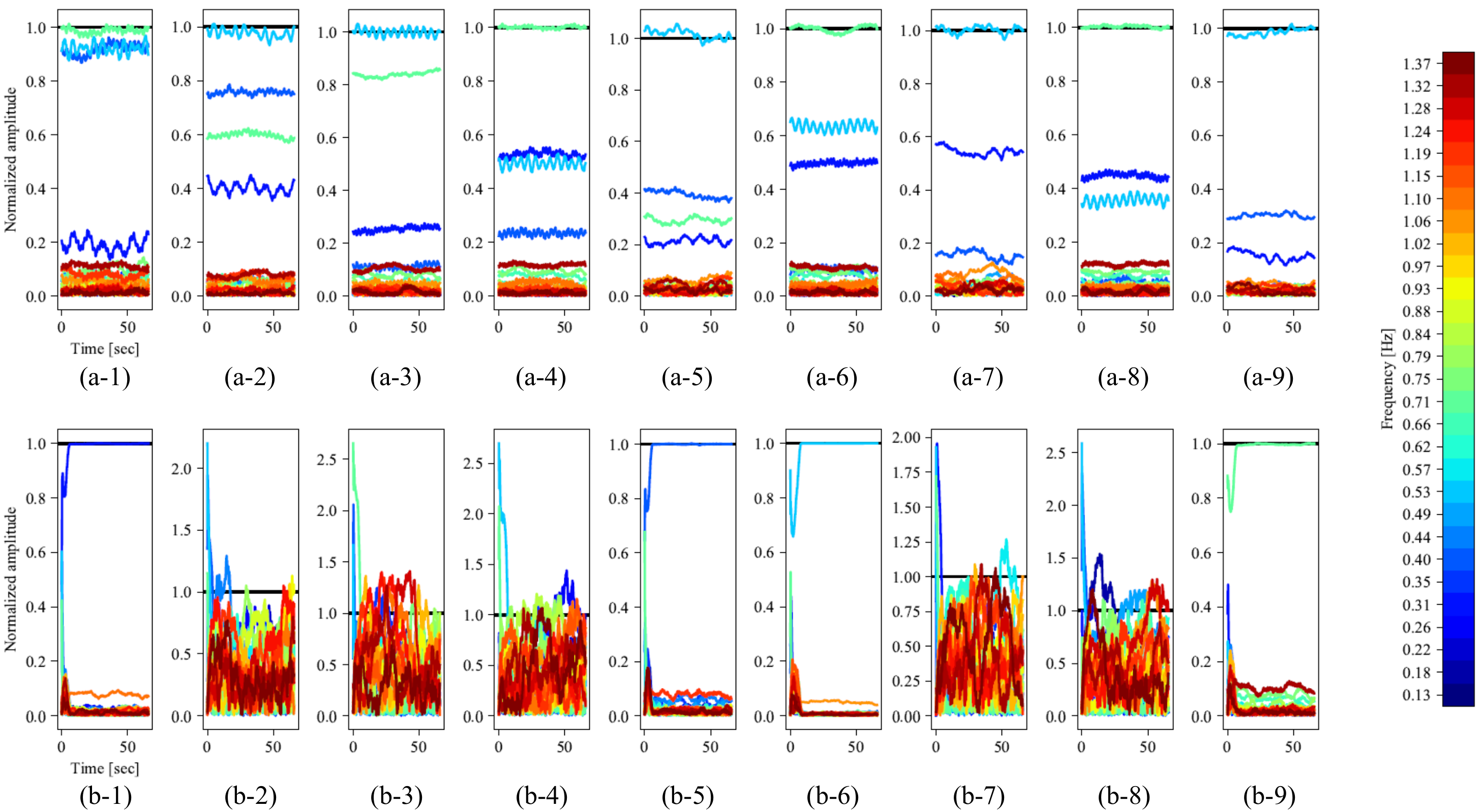}
  \caption{Temporal changes of the normalized signal magnitude at respective frequencies of (a-$\star$) mixed and (b-$\star$) separated signals obtained by HOT-ICA for the setting of three transmitting and three receiving antennas.}
  \label{fig:3x3specchanges}
\end{figure}

Fig.~\ref{fig:3x3specchanges} shows the temporal changes of the signal magnitude of respective frequencies in the spectra in Fig.~\ref{fig:3x3spec}.
The horizontal axis indicates time, and the color shows frequency.
We can see in Figs.~\ref{fig:3x3specchanges} (b-1), (b-5), (b-6) and (b-9) that the primary peak is determined well as a specific frequency in about 5 seconds, from which we can evaluate the speed of the HOT-ICA learning.
The secondary peak is also observed in Figs.~\ref{fig:3x3specchanges} (b-1), (b-5), (b-6) and (b-9).

\begin{figure}[H]
  \centering
  \includegraphics[width=1\linewidth]{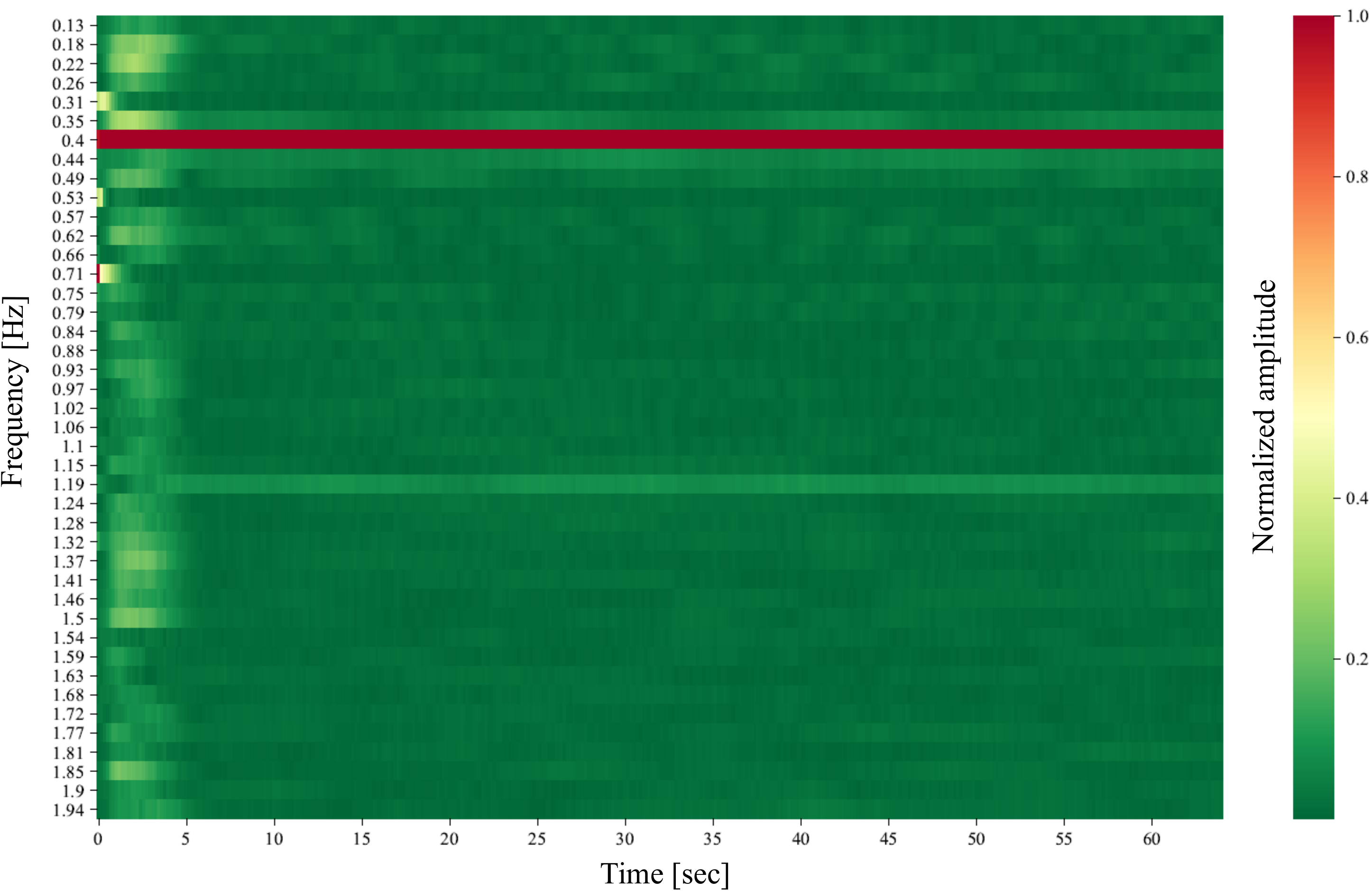}
  \caption{Spectrogram of separated signal for Target\,H1.}
  \label{fig:b-6spectrogram}
\end{figure}

\begin{figure}[H]
  \centering
  \includegraphics[width=1\linewidth]{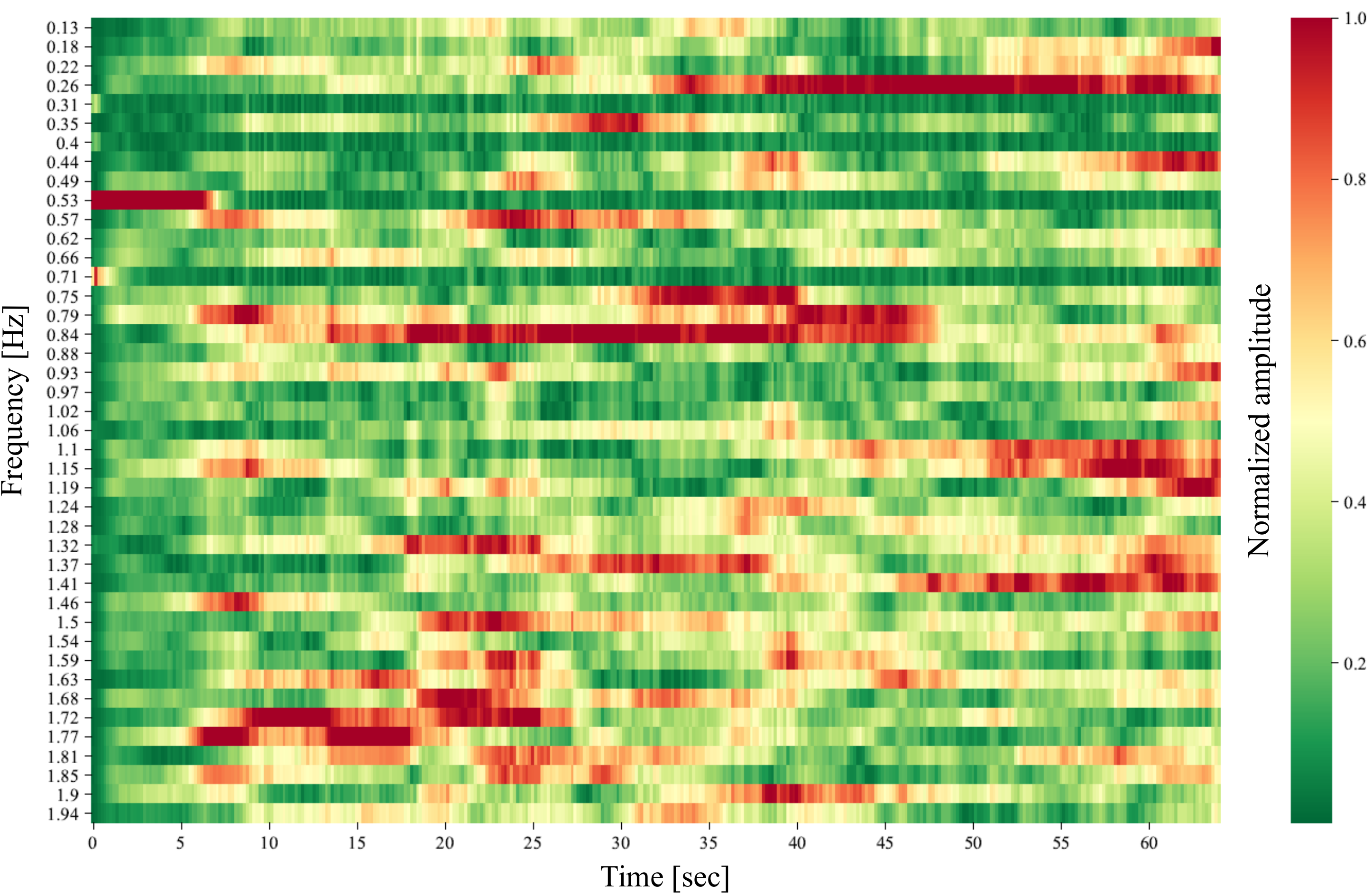}
  \caption{Spectrogram of separated noise.}
  \label{fig:b-9spectrogram}
\end{figure}

Fig.~\ref{fig:b-6spectrogram} shows the spectrogram of the signal separated for Target H1 corresponding to (b-5) in Figs.~\ref{fig:3x3spec} and \ref{fig:3x3specchanges}.
The vertical axis indicates frequency, the horizontal axis represents time, and the color shows the normalized signal magnitude.
We can see that the secondary peak appears as a light yellow-green straight line (1.19~Hz) as well as the primary peak (0.40~Hz) in red.
In Fig.~\ref{fig:b-9spectrogram}, the spectrogram shows the separated noise corresponding to (b-4) in Figs.~\ref{fig:3x3spec} and \ref{fig:3x3specchanges}.
In this graph, if we focus on the lines at respiration frequencies of 0.40~Hz, 0.31~Hz, 0.71~Hz and 0.53~Hz of targets H1, H2, H3 and H4, respectively, we find dark green straight lines there.
In other words, HOT-ICA strongly suppresses the target signals in the separated noise-only component.
From this result, we find that HOT-ICA works very effectively for the separation.

\subsection{Comparison of online CF-ICA and HOT-ICA without/with weights in separation tensor updates}
\label{subsec:comparative_experiment_cfhot}

\begin{figure}[H]
  \centering
  \includegraphics[width=0.8\linewidth]{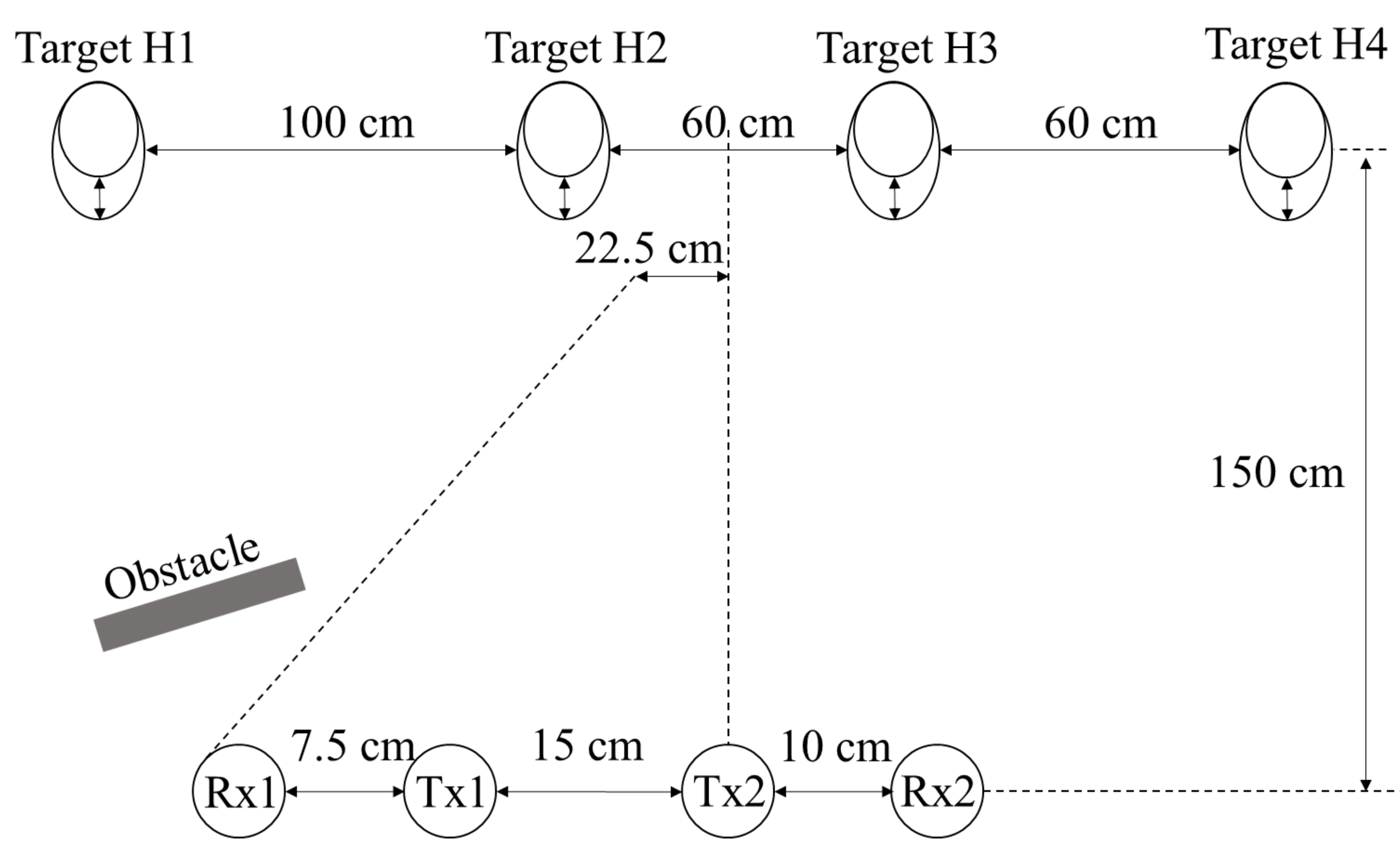}
  \caption{Placement of antennas, targets and an obstacle.}
  \label{fig:placement_debris}
\end{figure}

In Fig.~\ref{fig:placement_debris}, an obstacle is placed between Target H1 and Receiving antenna Rx1.
To simplify the situation, we assume a two transmitting-antenna and two receiving-antenna system.
The obstacle has two effects on the mixed signals $\mathbf{x}(t)$.
The first is that the obstacle gives an attenuation ($-50$~dB) to the original signal from Target H1 obtained by Receiving Antenna Rx1.
The second is that the signal received by Receiving Antenna Rx1 includes $+16$~dB noise compared to the noise at other receiving antennas because of its auto-gain control (AGC) in the following amplifier.

\subsubsection{Online CF-ICA}
\label{subsubsec:result_cf}

\begin{figure}[H]
  \centering
  \includegraphics[width=0.9\linewidth]{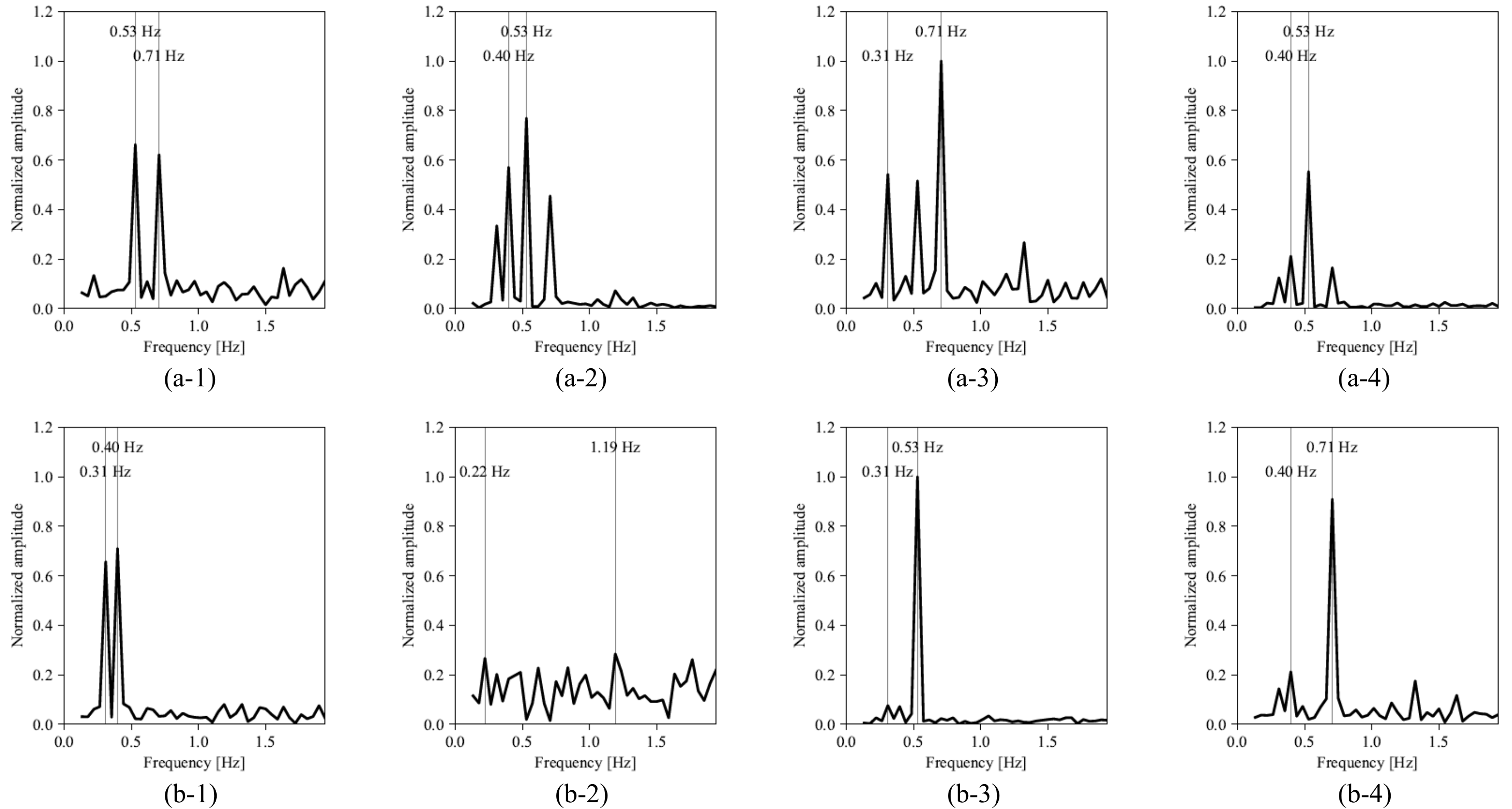}
  \caption{Spectra of (a-$\star$) mixed and (b-$\star$) separated signals using CF-ICA in the last time window for the setting of two transmitting and two receiving antennas in the environment with an obstacle.}
  \label{fig:obsta2x2cfspec}
\end{figure}
\begin{figure}[H]
  \centering
  \includegraphics[width=0.9\linewidth]{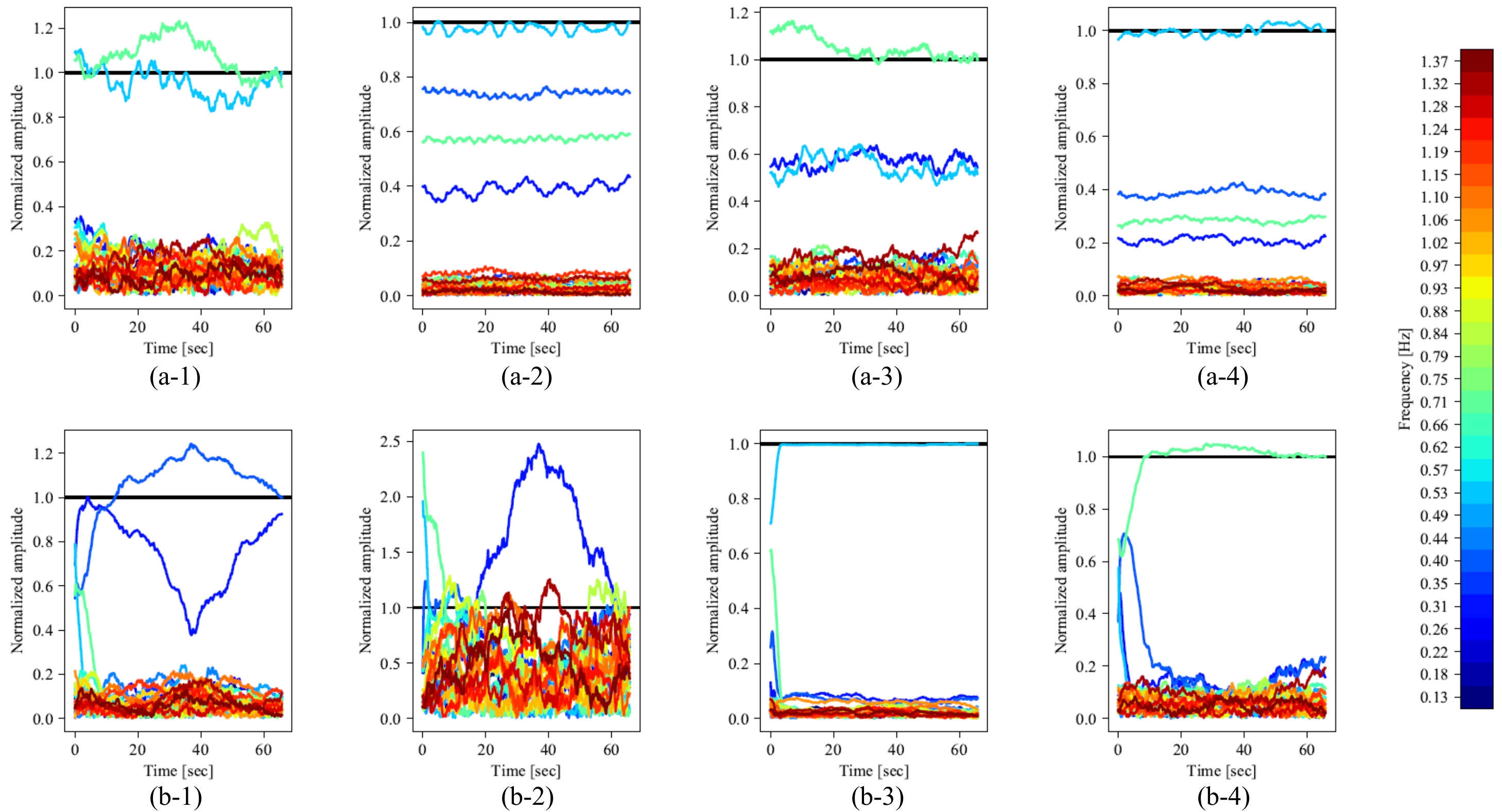}
  \caption{Temporal changes of the normalized signal magnitude at respective frequencies of (a-$\star$) mixed and (b-$\star$) separated signals using CF-ICA for the setting of two transmitting and two receiving antennas in the environment with an obstacle.}
  \label{fig:obsta2x2cfspecchanges}
\end{figure}

Fig.~\ref{fig:obsta2x2cfspec} shows the results of online CF-ICA processing when the obstacle affects the received signals.
The upper graphs in Fig.~\ref{fig:obsta2x2cfspec} show the received signal spectra $\mathbf{X}(\omega,\,T_d)$ at the last time window obtained by microwave transmitted by Tx1 and received by Rx1, denoted as (Tx1, Rx1) as well as those by (Tx1, Rx2), (Tx2, Rx1), and (Tx2, Rx2) in the order from left to right.
Large noise appears in the mixed signal spectra via Receiving Antenna Rx1 (see Figs.~\ref{fig:obsta2x2cfspec} (a-1) and (a-3)).
In the processed signals shown in Fig.~\ref{fig:obsta2x2cfspec} (b-1), the respiration frequencies of Targets H1 and H2 appear in the spectrum. That is, they are not separated well. On the otherhand, noise is dominant in Fig.~\ref{fig:obsta2x2cfspec} (b-2).

In Fig.~\ref{fig:obsta2x2cfspecchanges} (b-1), the respiration signal of Target H1 is decided as a peak, but that of Target H2 cannot be ignored.
In Fig.~\ref{fig:obsta2x2cfspecchanges} (b-2), the respiration signal of Target H2 sometimes becomes a peak during the learning, but noise is dominant.
We found that, when there is an obstacle, the separation learning using online CF-ICA becomes unstable.

\subsubsection{HOT-ICA without sensitivity control in separation tensor updates}
\label{subsubsec:result_noweight_hot}

\begin{figure}[H]
  \centering
  \includegraphics[width=0.9\linewidth]{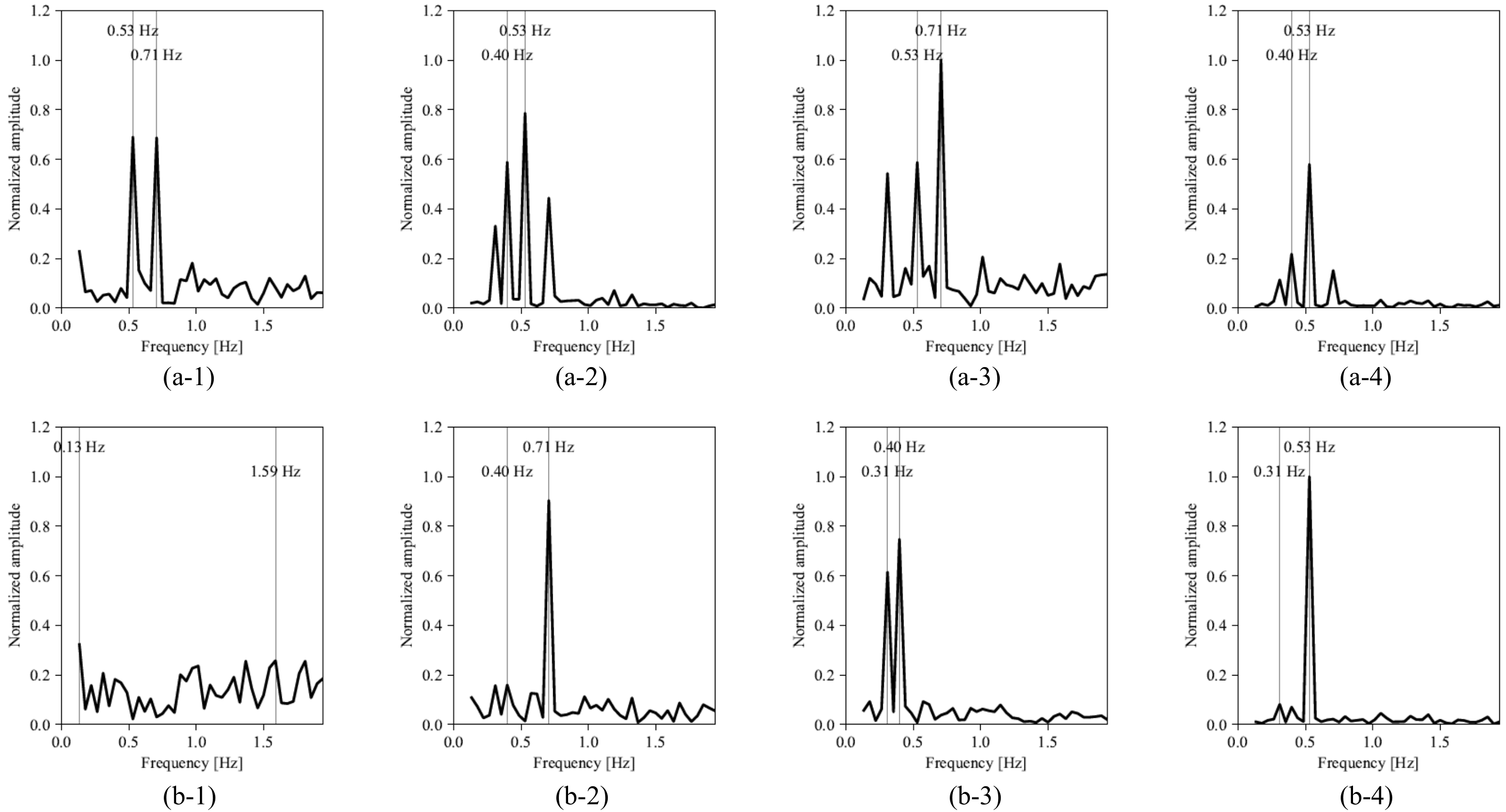}
  \caption{Spectra of (a-$\star$) mixed and (b-$\star$) separated signals using HOT-ICA without sensitivity control in separation tensor updates in the last time window for the setting of two transmitting and two receiving antennas in the environment with an obstacle.}
  \label{fig:obsta2x2spec}
\end{figure}
\begin{figure}[H]
  \centering
  \includegraphics[width=0.9\linewidth]{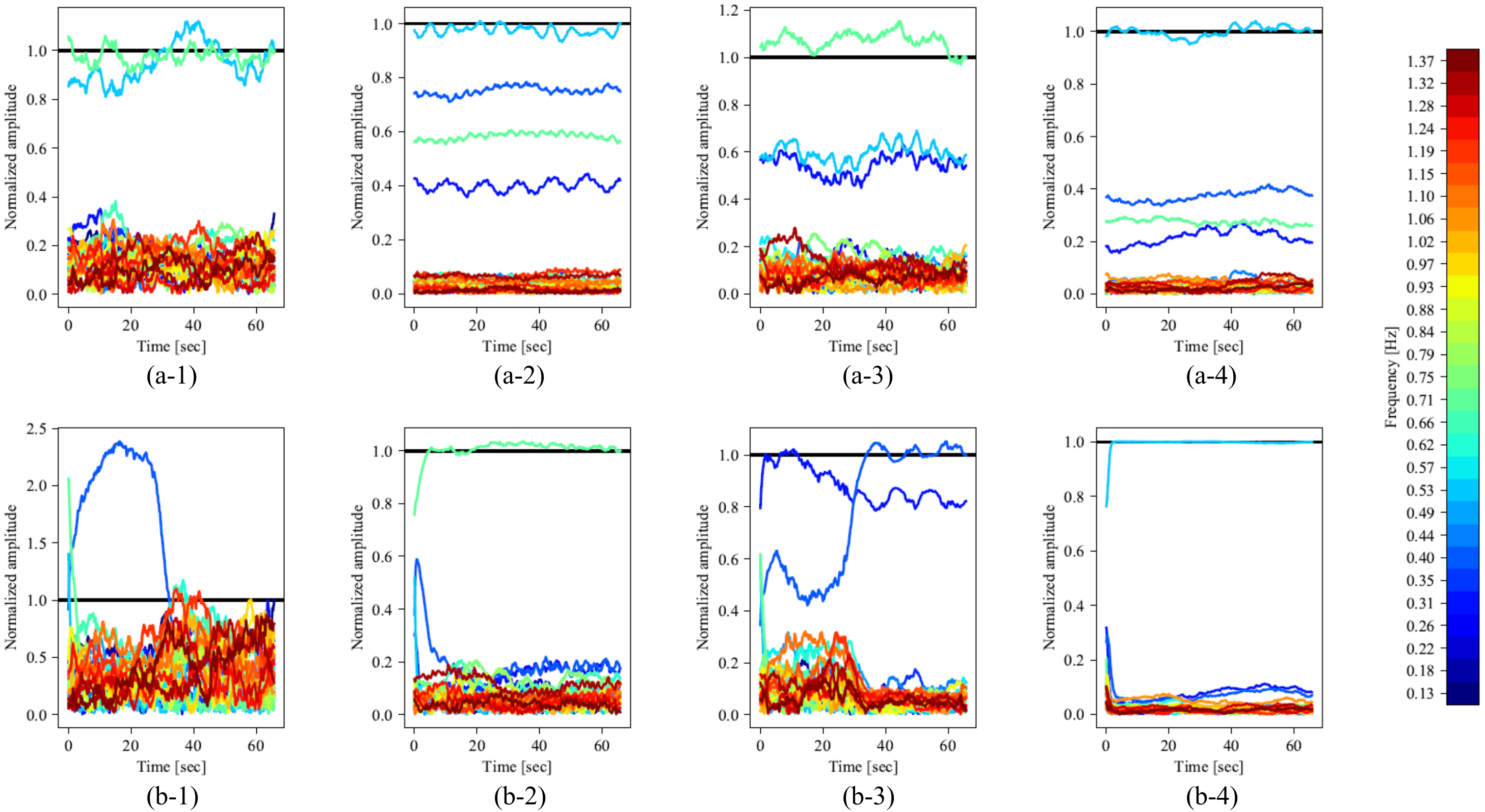}
  \caption{Temporal changes of the normalized signal magnitude at respective frequencies of (a-$\star$) mixed and (b-$\star$) separated signals using HOT-ICA without sensitivity control in separation tensor updates for the setting of two transmitting and two receiving antennas in the environment with an obstacle.}
  \label{fig:obsta2x2specchanges}
\end{figure}

As well as Section~\ref{subsubsec:result_cf}, we assume the same environment shown in Fig.~\ref{fig:placement_debris}.
The results of the HOT-ICA without sensitivity control in separation tensor updates are shown in Fig.~\ref{fig:obsta2x2spec}, where the obstacle affects the received signals.
The upper graphs in Fig.~\ref{fig:obsta2x2spec} show the mixed signal spectra $\mathbf{X}(\omega,\,T_d)$ at the last time window obtained by (Transmitting Antenna Tx1, Receiving Antenna Rx1), (Tx1, Rx2), (Tx2, Rx1), and (Tx2, Rx2) in the order from left to right.
Large noise appears again in the mixed signal spectra via Receiving Antenna Rx1 in Figs.~\ref{fig:obsta2x2spec} (a-1) and (a-3).
In Fig.~\ref{fig:obsta2x2spec} (b-3), the respiration frequencies of Targets H1 and H2 also appear in the same spectrum, without good separation. In Fig.~\ref{fig:obsta2x2spec} (b-1), noise is dominant.

In Fig.~\ref{fig:obsta2x2specchanges} (b-3), the respiration signal of Target H1 is recognized as a peak, but that of Target H2 is non-ignorable.
In Fig.~\ref{fig:obsta2x2specchanges} (b-1), the respiration signal of Target H2 sometimes becomes large, but noise is dominant.
When there is an obstacle, the separation learning using HOT-ICA without sensitivity control in separation tensor updates also becomes unstable.

\subsubsection{HOT-ICA with sensitivity control in the separation tensor updates}
\label{subsubsec:result_weight_hot}

\begin{figure}[H]
  \centering
  \includegraphics[width=0.9\linewidth]{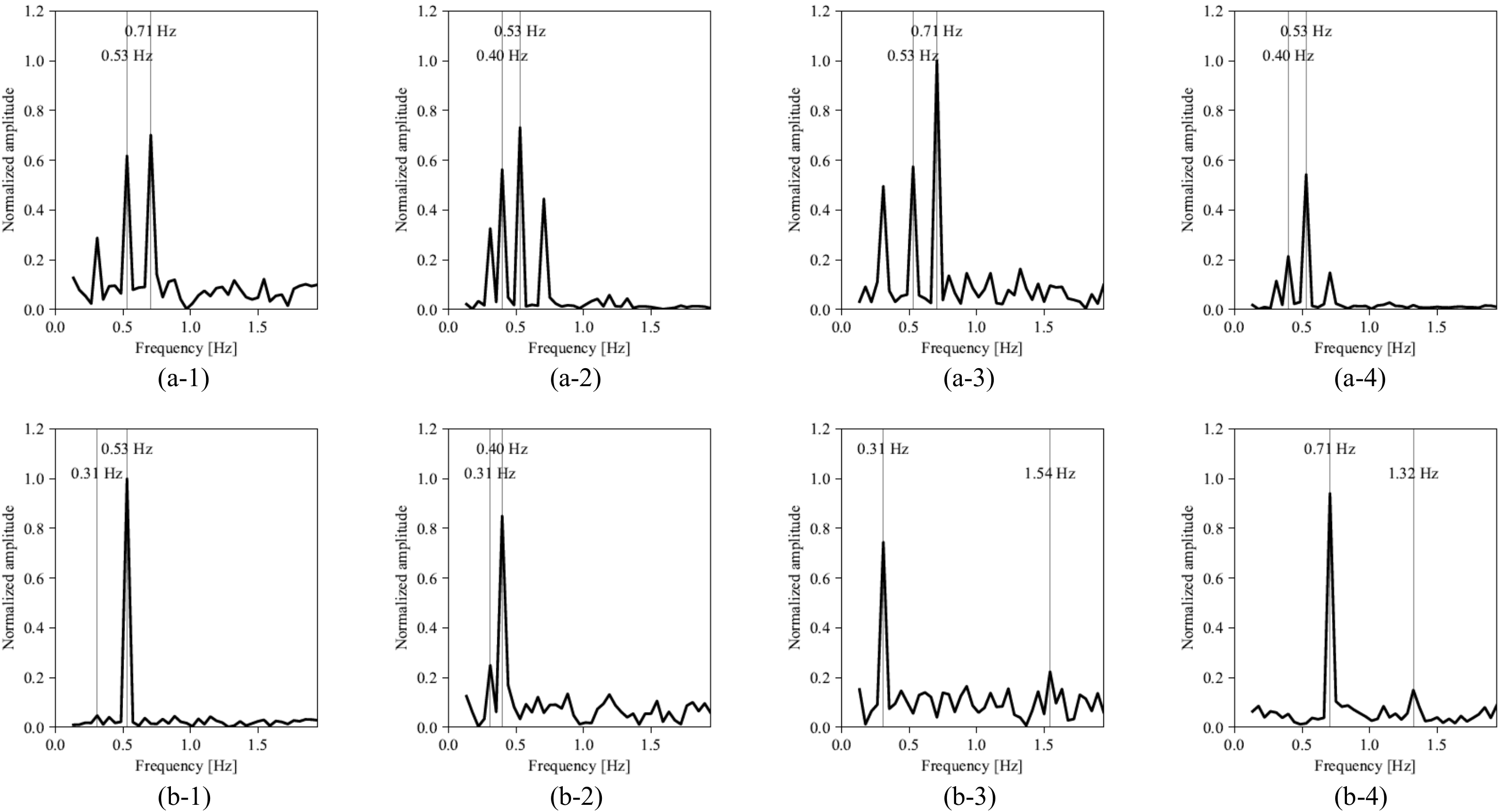}
  \caption{Spectra of (a-$\star$) mixed and (b-$\star$) separated signals using HOT-ICA with sensitivity control of $\eta_{\mathrm{Rx}1} = 0$ in the last time window for the setting of two transmitting and two receiving antennas in the environment with an obstacle.}
  \label{fig:rx1drop-obsta2x2spec}
\end{figure}

\begin{figure}[H]
  \centering
  \includegraphics[width=0.9\linewidth]{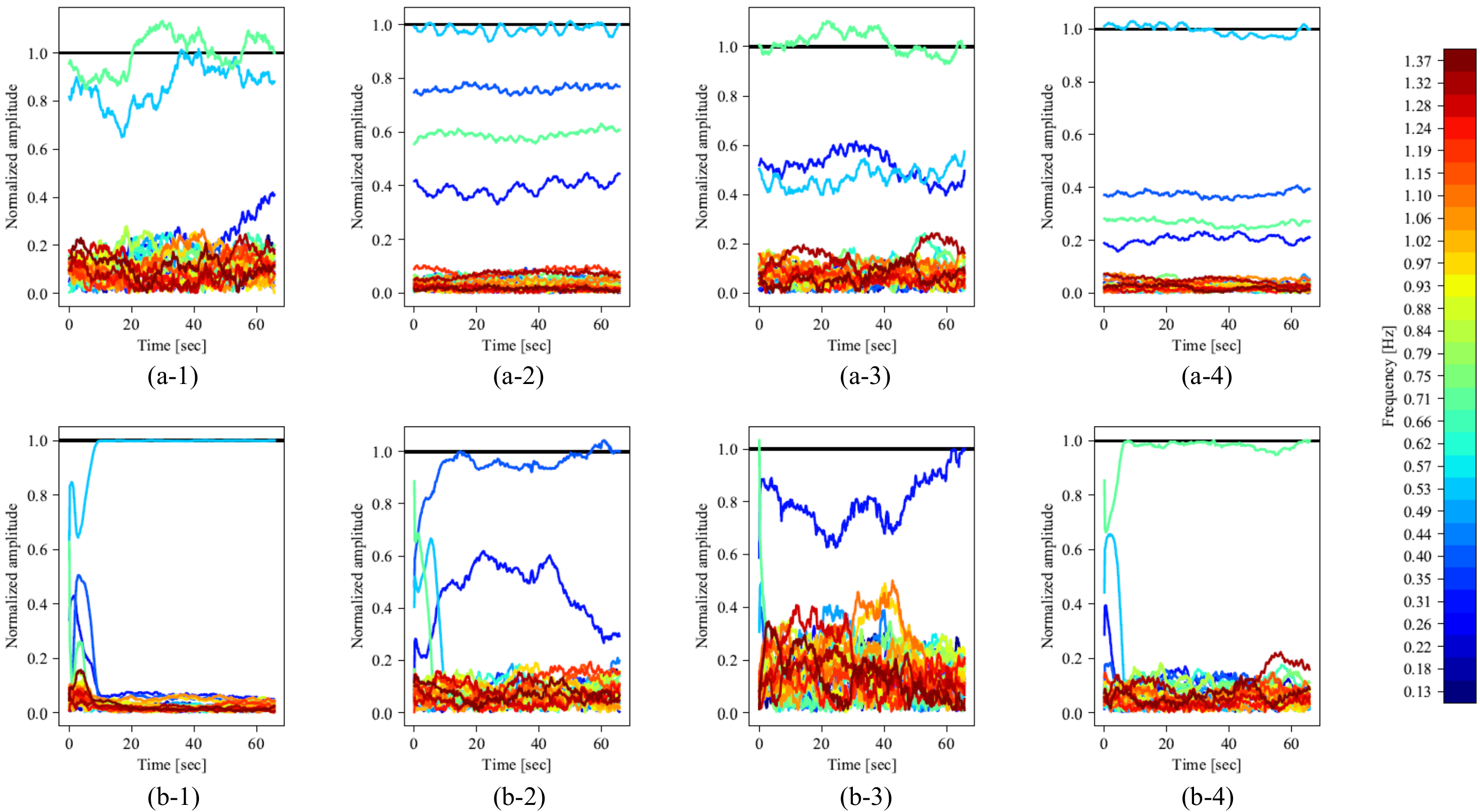}
  \caption{Temporal changes of the normalized signal magnitude at respective frequencies of (a-$\star$) mixed and (b-$\star$) separated signals using HOT-ICA with sensitivity control of $\eta_{\mathrm{Rx}1} = 0$ for the setting of two transmitting and two receiving antennas in the environment with an obstacle.}
  \label{fig:rx1drop-obsta2x2specchanges}
\end{figure}

Fig.~\ref{fig:rx1drop-obsta2x2spec} shows the results of the HOT-ICA with sensitivity control in relation to Receiving Antenna Rx1 in the same environment.
Specifically, we take $\eta_{\mathrm{Rx}1}=0$ in (\ref{eq:reduc_W_Rx1}). In practical systems, the reduction of mixed signals and/or the increase of noise are detectable in the front-end so that the sensitivity can be controled automatically.
In contrast with the results without the control in Fig.~\ref{fig:obsta2x2spec} (b-$\star$), each spectrum of the separated signals in Fig.~\ref{fig:rx1drop-obsta2x2spec} (b-$\star$) has a specific signal peak. Individual target signals are distributed over respective spectra separately.

We can see in Fig.~\ref{fig:rx1drop-obsta2x2specchanges} (b-$\star$) that the spectrum after separation finally determines a specific peak frequency.
Compared to Fig.~\ref{fig:3x3specchanges} (b-$\star$) including no obstacle, the obstruction delays the completion of separation learning. However, unlike Fig.~\ref{fig:obsta2x2specchanges} (b-1), the target signals are separated successfully without domination of noise in Fig.~\ref{fig:rx1drop-obsta2x2specchanges} (b-$\star$).
Thus, we have found the effectiveness of HOT-ICA with the sensitivity control related to the signal decrease and/or the noise increase caused by obstacles in the propagation paths.

HOT-ICA can include the control of sensitivity to respective components of the learning weight tensor $\underline{\underline{\mathbf{W}}}$ (fourth-order tensor in the above case).
In other words, it realizes direct control of the parameters in the learning dynamics.
This successful increase of robustness reveals the significance of keeping the categorization of the data in HOT-ICA.

\section{Conclusion}
\label{s:conclusion}

This paper proposed HOT-ICA. It is a new signal-separation method suitable for categorized data sets obtained by the measurement for human respiration and heartbeat employing a CW MIMO Doppler radar.
A numerical-physical experiment demonstrated that the HOT-ICA shows a good performance in separating target signals and noise online.
In addition, we set an obstacle in the environment, which causes the attenuation of target signals and the increase of noise. We compared the separation performances between with and without the sensitivity control in the separation tensor updates.
As a result, we found that the HOT-ICA with the sensitivity control is more robust to the obstacle occurrence than that without the control in signal-source separation learning, which leads to more flexible observation in various measurement situations. 
This robustness is achieved by HOT-ICA's signal processing dynamics that utilizes the nature of high-dimensional tensor structure which represents the categories in the data.

\section*{Acknowledgement}
The authors thanks Takahiro Nakanishi for his help in the experiments.


\bibliography{vital,source_separation_r,others,mica,topublish}

\end{document}